\documentclass[floatfix,amsmath,amssymb,aps,prc,twocolumn,superscriptaddress,showpacs,preprintnumbers,nofootinbib]{revtex4-1}

\usepackage{graphicx,color}
\usepackage{multirow}
\usepackage{times}
\usepackage{bm}
\usepackage{epstopdf}
\usepackage{enumerate}
\usepackage{ulem}
\usepackage{hyperref}
\usepackage{physics}

\newcommand{\GeV}{\textrm{GeV}}

\newcommand{\comment}[1]{}

\renewcommand\sout{\bgroup \color{red} \ULdepth=-.5ex \ULset}

\newcommand{\ssL}{{\scriptscriptstyle \Lambda}}
\graphicspath{{./Figs/}}

\newcommand{\sk}{\rho}
\begin{document}

\title{
Directed flow of $\Lambda$ in high-energy heavy-ion collisions
and $\Lambda$ potential in dense nuclear matter
}
\author{Yasushi Nara}
\affiliation{
Akita International University, Yuwa, Akita-city 010-1292, Japan}
\author{Asanosuke Jinno}
\affiliation{
Department of Physics, Faculty of Science,
Kyoto University, Kyoto 606-8502, Japan}

\author{Koichi Murase}
\affiliation{Yukawa Institute for Theoretical Physics,
Kyoto University, Kyoto 606-8502, Japan}

\author{Akira Ohnishi}
\affiliation{Yukawa Institute for Theoretical Physics,
Kyoto University, Kyoto 606-8502, Japan}

\date{\today}

\preprint{YITP-22-81, KUNS-2935}
\begin{abstract}
We investigate the sensitivity of the $\Lambda$ directed flow to the $\Lambda$ potential
in midcentral Au + Au collisions at $\sqrt{s_{NN}}\approx3.0$--$30$ GeV\@.
The $\Lambda$ potential obtained from the chiral effective field theory ($\chi$EFT) is used in a microscopic
transport model, a vector version of relativistic quantum molecular dynamics.
We find that
the density-dependent $\Lambda$ potentials, obtained from the $\chi$EFT
assuming weak momentum dependence of the potential,
reproduce the rapidity and the beam-energy dependence of the $\Lambda$ directed flow measured by the STAR collaboration
in the beam energy scan program.
Although the $\Lambda$ directed flow is insensitive to 
the density dependence of the potential, it is susceptible to the momentum dependence.
We also show that
a hydrodynamics picture based on the blast-wave model
predicts
a similarity of the proton, $\Lambda$, and $\Xi$ directed flows,
but the directed flow of $\Omega$ baryons slightly deviates from other baryons.
We also show that the quark coalescence predicts
different rapidity dependence of the directed flows
for hyperons.
These investigations suggest that measurements of a wide range of the rapidity dependence of the directed flow of hyperons
may provide important information about the properties of hot and dense matter created in high-energy heavy-ion collisions.

\end{abstract}

\maketitle

\section{Introduction}

The determination of the nuclear equations of state (EoS) is one of the most important problems in various fields, in which the properties of strongly
interacting quantum chromodynamics (QCD) matter plays an essential role, such as the dynamics of heavy-ion collisions~\cite{Danielewicz:2002pu}
or the structure and evolution of neutron stars~\cite{Oertel:2016bki}.

Anisotropic collective flows generated in the high-energy heavy-ion collision~\cite{Poskanzer:1998yz}
have been extensively investigated to extract the EoS of the dense QCD matter in a wide range of the incident energies.
The first Fourier coefficient 
of the distribution of the azimuth of the hadron momentum
($\phi$) relative to the reaction plane ($\Phi$)
is called the directed flow, $v_1=\langle \cos(\phi-\Phi)\rangle$.
It is predicted that the slope $dv_1/dy$ with respect to the rapidity ($y$) near midrapidity may show evidence of a first-order phase transition
due to softening of the EoS in the vicinity of the transition%
~\cite{Rischke:1995pe,Brachmann:1999xt,Csernai1999,Li:1998ze}.
The beam-energy dependence of the slopes for identified hadrons were measured by the STAR beam energy dcan (BES) programs%
~\cite{STAR2014,STAR2016,STAR:2017okv,STAR:2020dav,STAR:2021yiu},
where the transition of the positive slope
to the negative slope was discovered for both protons and $\Lambda$s
at $\sqrt{s_{NN}}\approx10$ GeV\@.
However, theoretical models with a first-order phase transition
predict this transition point at much lower beam energies%
~\cite{Steinheimer:2014pfa,Konchakovski:2014gda,Ivanov:2014ioa,Nara:2016hbg}.
A remarkable finding is that
almost identical directed flows are observed for protons and $\Lambda$s.
Many questions arise from the above energy dependence of the directed-flow slope of protons and $\Lambda$s:
Is this an indication of the transition from hadronic matter
to quark matter? Is this also an indication of the formation
of thermalized matter?
Can we extract information about the $\Lambda$ potential in a dense matter?

The proton directed flow in the collision energy range $2~\GeV <\sqrt{s_{NN}}<20~\GeV$
was shown to be explained by a transport model~\cite{JAMRQMDv}
with a purely hadronic EoS\@.
A repulsive EoS contributes positively to the slope
in the early stage (compression stage) of the collision,
while the tilted ellipsoid of the matter geometry contributes negatively in the late stage
(expansion stage).
The sum of these contributions causes the $dv_1/dy$ sign change at
$\sqrt{s_{NN}} \simeq 10~\GeV$~\cite{STAR2014,STAR2016,STAR:2017okv}
because the compression stages become shorter and the expansion stages get longer as the beam energy increases.
The transition point is sensitive to the interaction.
When only the Boltzmann-type two-body collisions are included,
which corresponds to the EoS of a free hadronic resonance gas,
interactions at the expansion stage are rather weak, and
the transition point shifts to higher beam energies%
~\cite{Konchakovski:2014gda}.
However, when mean-field interactions are added to simulate
interacting hadronic matter, transition point shifts to lower beam energies~\cite{JAMRQMDv}.

Thus, we expect that the $\Lambda$ directed flow is also sensitive
to the $\Lambda$ potential in highly dense matter,
and 
the $\Lambda$ single-particle potential $U_\Lambda$ may be constrained by the $\Lambda$ flow.
Specifically, $U_\Lambda$
from the chiral effective field theory ($\chi$EFT)
is promising~\cite{GKW2020}.
This is complementary to the precision hypernuclear spectroscopy.
The $\chi$EFT predicts a very strong repulsion at high densities
by the $\Lambda NN$ three-body interactions~\cite{GKW2020}.
It is argued that this strong repulsive potential
may solve the hyperon puzzle of the neutron stars, which
is one of the primary problems in the neutron-star physics~\cite{HyperonPuzzle}.
$\Lambda$ baryons are expected to appear in the neutron-star matter
at about two to four times the normal nuclear matter density
when only the two-body interactions based on hypernuclear data are used.
However, hyperons, which soften the EoS, make it difficult
to explain the existence of two-solar-mass neutron stars.
One of the possibilities to avoid the hyperon puzzle
is to suppress $\Lambda$ in dense nuclear matter by, for example
a strong repulsive potential 
for $\Lambda$ in dense nuclear matter.

In this work, we will investigate the effects of the $\Lambda$ potential
on the directed flow by using a microscopic transport model.
One of our purposes is to examine whether 
the strong repulsive potential predicted by $\chi$EFT can explain 
the data of the $\Lambda$ directed flow from the STAR collaboration.
We are also interested in whether the directed-flow data can constrain
$U_\Lambda$ at high densities.
A blast-wave analysis with rapidity-dependent flows on hyperons
will be performed as well.

To extract information about the properties of hot and dense nuclear matter from heavy-ion collisions, we have to compare theoretical predictions with experimental data.
Nonequilibrium transport theoretical approaches
such as Boltzmann-Uhling-Uhlenbeck~\cite{Bertsch:1988ik,Cassing:1990dr,Ko:1987gp,Blaettel:1993uz,GiBUU} or
quantum molecular dynamics (QMD)~\cite{Aichelin:1991xy,RQMD1989,UrQMD1,Isse:2005nk,Aichelin:2019tnk} type models have been successfully used to understand the data. The microscopic transport models are suitable to extract the EoS from heavy-ion data~\cite{Danielewicz:2002pu,Li:1998ze,Danielewicz:1998vz,Rai:1999hz,Hillmann:2018nmd,Oliinychenko:2022uvy,Steinheimer:2022gqb}.
The microscopic transport models combine
the mean-field interaction and the Boltzmann-type collision term
at a semiclassical level.
A recent compilation of microscopic transport models
can be found in Ref.~\cite{TMEP:2022xjg}.
In this paper, we study the directed flow of $\Lambda$
by using the transport model, a Lorentz-vector version of relativistic quantum molecular dynamics (RQMDv)~\cite{JAMRQMDv} in the JAM event generator,
and verify the $\Lambda$ potential at high densities.
This will be a first verification of the strong
repulsion in the $\Lambda$ potential using heavy-ion data.

This paper is organized as follows.
In Sec.~\ref{sec:lambdapot}, we introduce the $\Lambda$ potential
to be used in the transport model.
Section~\ref{sec:rqmd} briefly summarizes the microscopic transport
model RQMDv.
In Sec.~\ref{sec:result}, we compare the beam-energy dependence as well as
the rapidity dependence of the directed flow of protons
and $\Lambda$s with the STAR data. We also discuss the collision
dynamics of how the $\Lambda$ directed flow is generated,
which is slightly different from the proton case.
As a complementary study, we present results in
a hydrodynamic picture by using the blast-wave model~\cite{Dobler:1999ju} in Sec.~\ref{sec:bwm}.
The conclusion and outlook are given in Sec.~\ref{sec:summary}.

\section{$\Lambda$ potential from chiral effective field theory}
\label{sec:lambdapot}

The $\Lambda$ potential at finite density has been studied extensively 
in nonrelativistic~\cite{LambdaPot-NonRel} and relativistic~\cite{LambdaPot-Rel} models.
These theories describe the $\Lambda$ separation energies of various hypernuclei
and have been applied to the EoS of the neutron-star matter.
For example, by using the spin-flavor SU(6) symmetry for the vector coupling, $g_{\omega\Lambda}/g_{\omega N}\simeq 2/3$,
one can fit the separation energy data of various hypernuclei by tuning the scalar coupling. Then one expects that dense matter EoS with hyperons can be predicted.
These simple treatments of hyperon potentials are found to fail in sustaining massive neutron stars~\cite{HyperonPuzzle},
which is now known as the hyperon puzzle.
To solve the hyperon puzzle, many ideas have been proposed~\cite{HyperonPuzzleSolution},
but most of these prescriptions contain additional parameters related to the three-baryon
interactions or the density dependence of the interactions,
which have not been constrained by the existing data.

One way to systematically describe many-body interactions
is to use the $\chi$EFT~\cite{chiralEFT}.
The $\chi$EFT is based on the chiral symmetry of massless QCD,
and finite quark mass and finite momentum effects can be introduced systematically
by introducing higher-order diagrams.
The $\chi$EFT with hyperons in the leading order (LO) and next-to-leading order (NLO) diagrams
has been given~\cite{Haidenbauer-NLO}, while the three-baryon interactions appear
in the next-to-next-to-leading order (NNLO) diagrams for the octet baryons.
A part of the NNLO diagrams relevant to the $\Lambda NN$ three-baryon force is 
included in the calculation of the $\Lambda$ potential in the nuclear matter~\cite{Kohno:2018gby}.
With decuplet baryons, by contrast,
a part of three-baryon diagrams can be evaluated
with the low-energy constants (LECs) determined in the LO and NLO diagrams~\cite{Decuplet}.

\begin{figure}[tbhp]
\includegraphics[width=11cm]{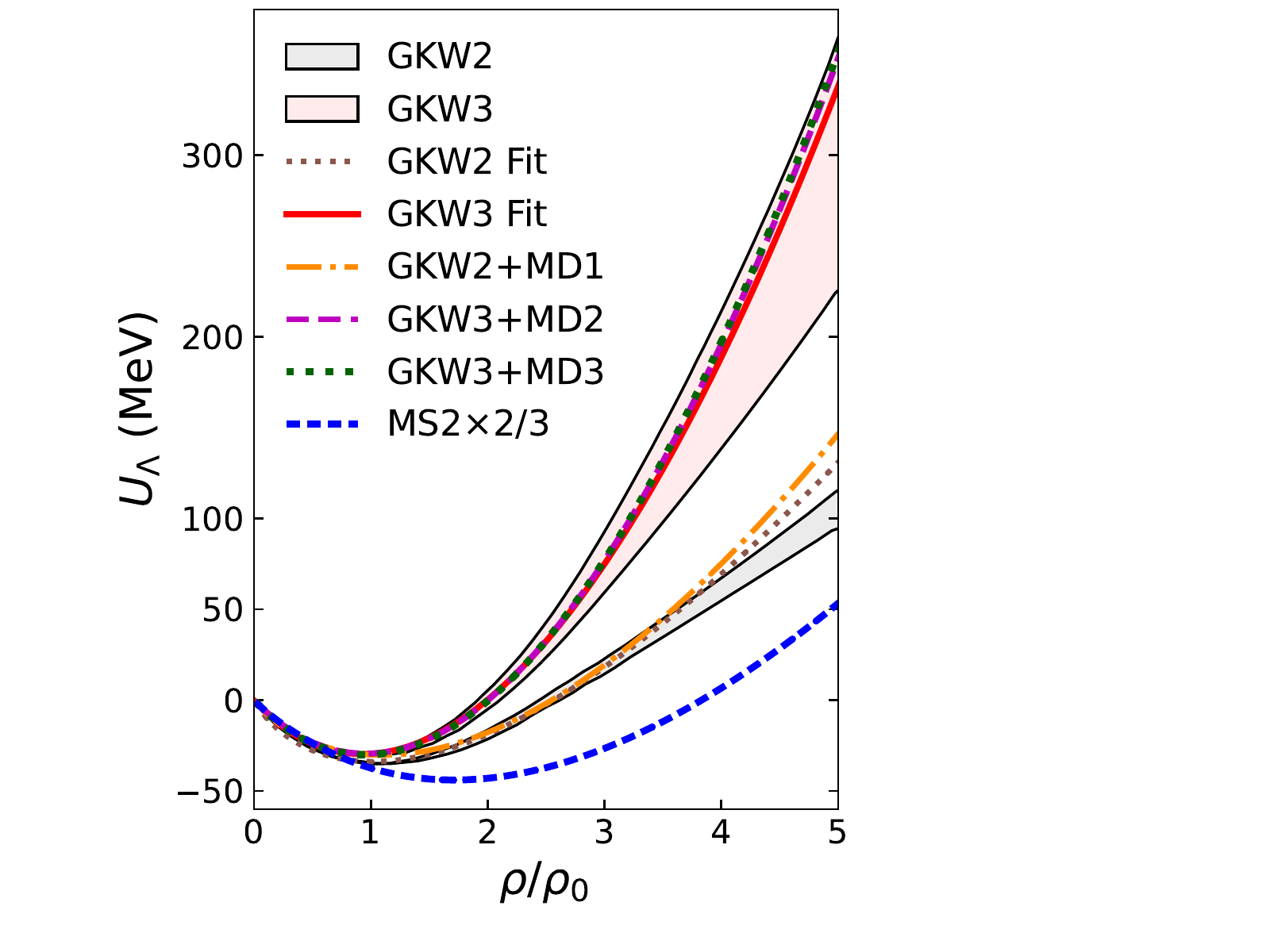}
\caption{Normalized baryon density dependence of the single-particle potentials for $\Lambda$.
GKW2 presents the results of the $\Lambda$ single-particle potential
with two-body interactions, while GKW3 is obtained
by the inclusion of three-body interactions.
Dash-dotted, bold-dashed, and bold-dotted lines correspond to the momentum-dependent potentials of GKW2+MD1, GKW3+MD2, and GKW3+MD3, respectively.
The dotted line corresponds to the nucleon single-particle potential
multiplied by two-thirds motivated by the light quark counting.
}
\label{fig:DensePot}
\end{figure}

As shown in Fig.~\ref{fig:DensePot},
the $\Lambda$ potential in nuclear matter at the zero momentum of $\Lambda$
is computed in the framework of Br\"uckner-Hartree-Fock theory
by using the $\chi$EFT in Ref.~\cite{GKW2020},
where the diagrams relevant to the three-body forces
are assumed to be saturated by the decuplet baryon propagation~\cite{Decuplet}.
The ultraviolet momentum cutoff in the $\chi$EFT
was chosen to be $\lambda=500~\textrm{MeV}/c$
in \cite{GKW2020}.
We have fitted 
the density dependence of the single-particle $\Lambda$ potential $U_\Lambda$ from $\chi$EFT
by using
the Fermi momentum expansion of the potential~\cite{Tews2017},
\begin{equation}
U_{\sk\Lambda}(\rho)=au+bu^{4/3}+cu^{5/3}\,,
\label{eq:UL}
\end{equation}
where $u=\rho/\rho_0$ is the nucleon density normalized by the saturation density
$\rho_0=0.168$ fm$^{-3}$.
The total $\Lambda$ potential is given by
\begin{equation}
   V_{\Lambda N} = \int d\bm{r} \int d\rho_\ssL(\bm{r})\, U_{\sk\Lambda}(\rho(\bm{r}))
    =\int d\bm{r}\, \rho_\ssL(\bm{r}) U_{\sk\Lambda}(\rho(\bm{r})).
    \label{eq:VL}
\end{equation}
Thus, single particle potential defined by the derivative
with respect to the phase-space distribution function $f(x,p)$,
$U\equiv \delta V/\delta f$ for $\Lambda$,
is $U_{\sk\Lambda}$ itself.
The $\Lambda\Lambda$ interaction is not included
in the present work, since its effect is small
for the directed flow
in the colliding-energy range under consideration.
The nucleon potential is taken from Ref.~\cite{JAMRQMDv}.

In the calculation of collective flows,
the momentum dependence of the potential is known to be important. 
We introduce the Lorentz-vector--type momentum-dependent potential~\cite{RBUUp} for $\Lambda$
\begin{equation}
 U_{m\Lambda}^\mu (\rho, p)= 
     \frac{C}{\rho_0}
     \int d^3{p}'
     \frac{p^{*'\mu}}{p^{*0'}}
     \frac{f(x,p')}{1+[(\bm{p}-\bm{p}')/\mu]^2},
     \label{eq:mdvec}
\end{equation}
where $p^{*\mu} = p^\mu - U^\mu$
and $p^{*0}=\sqrt{m_N^2+\bm{p}^{*2}}$ from the mass-shell constraint.
Some expressions for the momentum-dependent vector potential $U^\mu$ in the cold nuclear matter are summarized
in the Appendix.
We note that two-range Lorentzian-type momentum-dependent potential
does not change the accuracy of the fitting.
The potential parameters $a+b+c$, $C$, and $\mu$
are fixed by fitting the optical potential $U_\mathrm{opt}$
to the momentum dependence of the $\Lambda$ single-particle potential~\cite{Kohno:2018gby}.
The optical potential $U_\mathrm{opt}$ is defined by
the difference of the single-particle energy and the kinetic
energy [see Eq.~\eqref{eq:VecOptPot} in the Appendix]. 
The parameters $a$, $b$, and $c$ in the density-dependent part of the potential are then obtained under the constraint of $a+b+c$,
which is already determined by the momentum-dependent part
of the potential.
The potential in Ref.~\cite{Kohno:2018gby}
is obtained by the decuplet saturation, but incorporates fewer diagrams than Ref.~\cite{GKW2020}.

We consider three types of momentum dependence.
MD1 and MD2 momentum-dependent potentials are obtained by
fitting the $\chi$EFT result of Ref.~\cite{Kohno:2018gby}
up to the momentum of 2.5 fm$^{-1} \simeq \lambda$ (cutoff)
at the normal nuclear density
assuming the range parameter $\mu=3.23$ fm$^{-1}$.
This range parameter is motivated by 
the nucleon potential in Ref.~\cite{JAMRQMDv}.
To construct a weaker momentum-dependent potential, MD3 momentum-dependent potential 
is obtained by solving the system of equations for $C$, $\mu$, and $a+b+c$:
\begin{align}
    U_{\mathrm{opt}}(\rho_0, p=0 \mathrm{~\textrm{fm}^{-1}}) &= -30 \textrm{~MeV}, \\
    U_{\mathrm{opt}}(\rho_0, p=1 \mathrm{~\textrm{fm}^{-1}}) &= -23.4 \textrm{~MeV},
    \label{eq:1fm}\\
    U_{\mathrm{opt}}(\rho_0, p=1.7 {\textrm{~GeV}}) &= 0 \textrm{~MeV}.
\end{align}
Equation~\eqref{eq:1fm} is the condition to reproduce the 
results of Ref.~\cite{Kohno:2018gby} up to the momentum of 1 fm$^{-1}$,
which is around 40\% of the cutoff.

\begin{table*}[tbhp]
\caption{Parameter set of the single-particle $\Lambda$ potential $U_\Lambda$ in Eqs.~\eqref{eq:UL} and \eqref{eq:mdvec} and the Taylor coefficients for $U_\Lambda$ around $u=1$ in Eqs.~\eqref{eq:taylor1}, \eqref{eq:taylor2}, and \eqref{eq:taylor3}.}
\label{Tab:ULpars2}
\begin{tabular}{l|rrrrrrrr}
\hline
\hline
Model           & $a$ (MeV) 
                            & $b$ (MeV) 
                                        & $c$ (MeV)
                                                   & $C$ (MeV)
                                                   & $\mu$ (fm$^{-1}$)
                                                   & $J_\Lambda$ (MeV)
                                                   & $L_\Lambda$ (MeV)
                                                   & $K_\Lambda$ (MeV)\\
\hline
GKW2 (2-body)   &  $-154.9$ & $142.4$   & $-21.4$  & $-$ & $-$ & $-33.83$ & $-1.825$ & $356.0$\\
GKW3 (2+3-body) &  $-80.1$  & $0.16$    & $50.4$   & $-$ & $-$ & $-29.55$ & $12.34$ & $504.8$\\
\hline
GKW2+MD1      &  $3.54$   & $58.99$    & $1.911$ & $-104.3$ & $3.23$ & $-30.12$ & $-10.18$ & $334.3$\\
GKW3+MD2      &  $58.8$   & $-42.60$  & $59.71$  & $-116.2$ & $3.23$ & $-29.41$ & $7.916$ & $515.0$ \\
GKW3+MD3      & $-1.072$  & $-69.58$  & $71.58$  & $-54.993$& $1.124$ & $-30.00$ & $8.740$ & $517.2$ \\
\hline
\hline
\end{tabular}
\label{table:lambda_eos}
\end{table*}

The results of fitting parameters are
summarized in Table~\ref{fig:DensePot}.
The Taylor coefficients for $U_\Lambda=U_{\rho\Lambda}+U_{m\Lambda}^0$ around the normal nuclear density,
\begin{align}
    J_\Lambda&=U_\Lambda(u=1), \label{eq:taylor1}\\
    L_\Lambda&=3\rho \dfrac{\partial U_\Lambda}{\partial \rho}\Bigr|_{\rho_0}, \label{eq:taylor2}\\
    K_\Lambda&=9\rho^2 \dfrac{\partial^2 U_\Lambda}{\partial \rho^2}\Bigr|_{\rho_0}, \label{eq:taylor3}
\end{align}
are also summarized in Table~\ref{fig:DensePot}.

We show the density dependence of 
the $\Lambda$ single-particle potential in Fig.~\ref{fig:DensePot}.
We show the results for the cases
where only the two-body interaction is included (GKW2)
and the three-body interactions are also included (GKW3).
GKW3 predicts much stronger density dependence than GKW2.
Since the cutoff of 500 MeV$/c$ is adopted
and the Br\"uckner-Hartree-Fock calculation using the $\chi$EFT interaction
is found to be unstable at $\rho/\rho_0\geq 3.5$ in \cite{GKW2020},
we have fitted to the upper and lower bound curves in the range
$\rho/\rho_0\leq 3$, and the average of these results is shown 
as fit results.
We note that the GKW3 potential suppresses the appearance of $\Lambda$ hyperons in neutron matters,
and thus, one may avoid the softening of the equation of state.
As a comparison, two-thirds of the nucleon single-particle
potential is plotted in Fig.~\ref{fig:DensePot},
which is often assumed in the transport models as a simple
recipe for implementing $\Lambda$ potential.

\begin{figure}[tbhp]
\includegraphics[width=8cm]{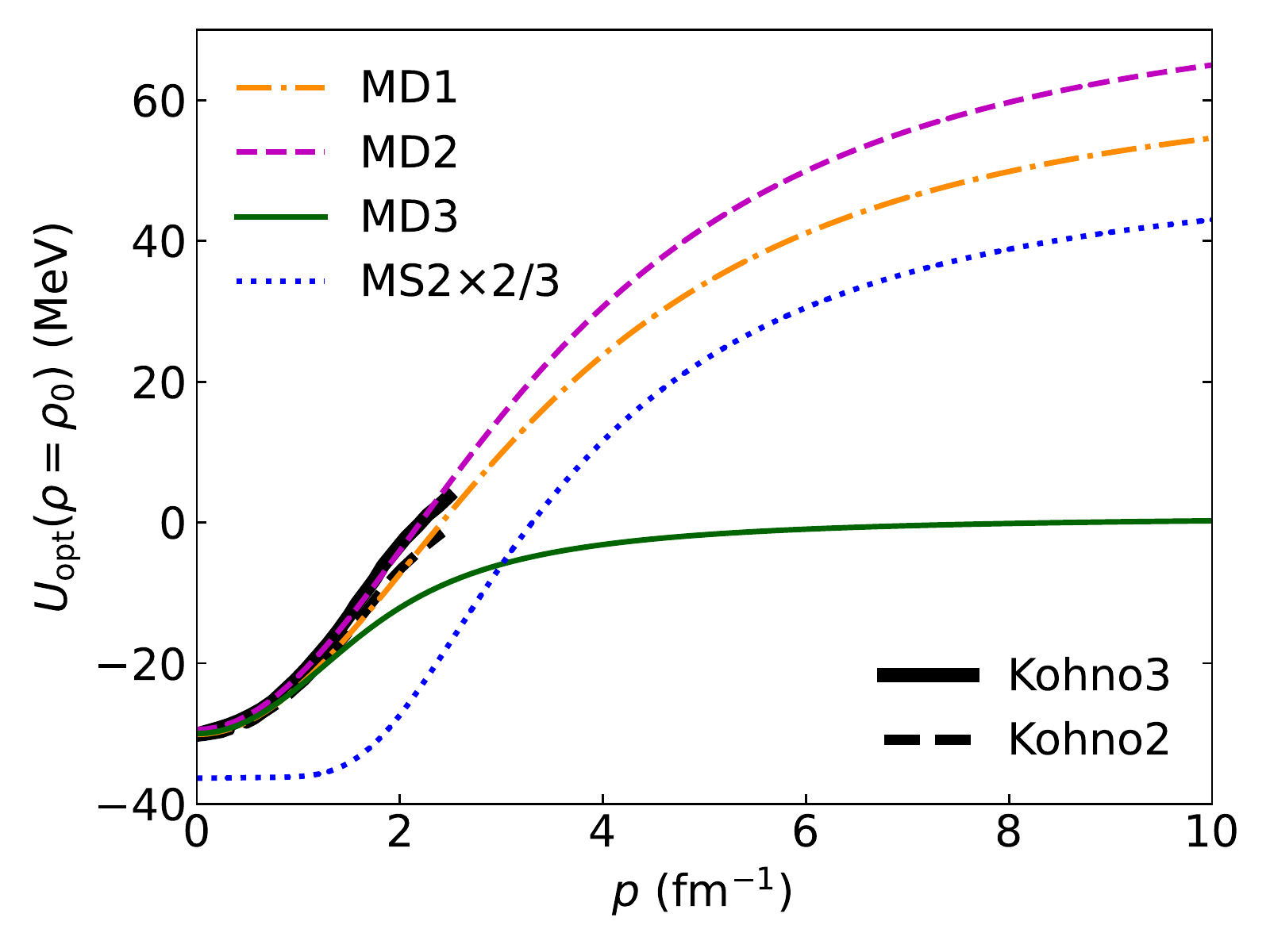}
\caption{Momentum dependence of the $\Lambda$ optical potentials at the normal nuclear density. 
Dash-dotted, dashed, solid, and dotted lines
correspond to MD1, MD2, MD3, and two-thirds of nucleon potential
parametrizations, respectively.
}
\label{fig:MomPot}
\end{figure}

Figure~\ref{fig:MomPot} shows the optical potentials with the MD1, MD2, and MD3 parameter sets
by the dash-dotted, dashed, and solid lines, respectively.
For comparison, two-thirds of the nucleon optical potential ($\text{MS2}\times 2/3$)
is plotted by the dotted line, which is small
compared to the prediction by the $\chi$EFT theory.

In the calculation of the heavy-ion collisions,
we assume that all hyperons ($\Lambda^{(*)}, \Sigma^{(*)}, \Xi^{(*)}, \Omega$)
are assumed to feel the same potential as $\Lambda$,
and all nonstrange baryons ($N^{(*)}, \Delta^{(*)}$) are assumed to feel the same potential as $N$.
Then, the nucleon density ($\rho$) in Eqs.~\eqref{eq:UL}, \eqref{eq:VL}, and \eqref{eq:mdvec}
is replaced with the net nonstrange baryon density,
and the $\Lambda$ density ($\rho_\ssL$) is replaced with the net-hyperon density.
Different potentials for different hyperon species will be discussed elsewhere.

\section{Relativistic quantum molecular dynamics}
\label{sec:rqmd}

We implement the above-mentioned potentials
in the form of Lorentz-vector potential
in the relativistic quantum molecular dynamics (RQMDv) approach developed in Ref.~\cite{JAMRQMDv}.
The RQMDv equations of motion for the $i$-th particle
having the position $q^\mu_i$ and the momentum $p^\mu_i$%
~\cite{Nara:2019qfd,Nara:2020ztb} is given by
\begin{align}
\frac{dq^\mu_i}{dt} & =
   v^{*\mu}_i
     -\sum_{j} 
      v^{*\nu}_j
      \frac{\partial {V}_{j\nu}}{\partial p_{i\mu}},
        \nonumber\\
\frac{dp^\mu_i}{dt}
      &= \sum_{j} 
      v^{*\nu}_j
     \frac{\partial V_{j\nu}}{\partial q_{i\mu}},
\end{align}
where $v_i^{*\mu}=p_i^{*\mu}/p_i^{*0}$.
The density-dependent part of the vector potential is defined by using the baryon current%
~\cite{Danielewicz:1998pb,Sorensen:2020ygf}
\begin{equation}
V^\mu_{\sk i} = B_i\frac{V_{\sk i}(\rho_{Bi})}{\rho_{Bi}} J_i^\mu,
\label{eq:vectorpotential}
\end{equation}
where $V_{\sk i}$ is the density-dependent (Skyrme-type) potential for baryons.
In this work, we use the MS2 EoS in Ref.~\cite{JAMRQMDv} for non-strange baryons.
For $\Lambda$ and other hyperons, we use $U_{\sk\Lambda}$ in Eq.~\eqref{eq:UL}.
The invariant baryon density
$\rho_{Bi}=\sqrt{J_i^\mu J_{i\mu}}$
is obtained from the baryon current $J^\mu_i$
\begin{equation}
J^\mu_i=\sum_{j\neq i}B_j \frac{p^{\mu}_j}{p^0_j} \rho_{ij},
\label{eq:RQMDdensity}
\end{equation}
where the sum runs over all the non-strange baryons,
$B_j$ is the baryon number of the $j$-th particle,
and $\rho_{ij}$ is the so-called interaction density
(i.e., the overlap of density with another hadron wave packet)
\begin{equation}
 \rho_{ij}=\frac{\gamma_{j}}{(4\pi L)^{3/2}}\exp(q^2_{Rij}/4L)\,,
 \label{eq:twobody}
\end{equation}
where $\gamma_j=p_j^{0}/m_j$, and $q^2_{Rij}$ is the squared distance in the rest frame of the particles $j$,
\begin{equation}
 q^2_{R,ij} = (q_i - q_j)^2 - [(q_i - q_j)\cdot u _j]^2,\quad
 u_j = p_j/m_j.
 \label{eq:restframedistance}
\end{equation}
This is used in the relativistic Landau-Vlasov model~\cite{RLV}, in which
Gaussian-shaped test particles are used to solve the relativistic Boltzmann-Vlasov equation.
We use the following vector-type momentum-dependent one-particle potential for $\Lambda$:
\begin{equation}
 V_{mi}^\mu (p_{R,ij})=\frac{C}{\rho_0}
 \sum_{j\neq i}
     \frac{p_j^{\mu}}{p_j^{0}}
     \frac{\rho_{ij}} {1-[p_{R,ij}/\mu]^2}
     \label{eq:MVpotV}
\end{equation}
with the parameters determined in the previous section.
%
%
The two-body relative momentum squared, $p_{R,ij}^2$, in the rest frame of
the particle $j$
is used for the argument of the momentum-dependent potential.

\section{Numerical implementation}
\label{sec:model}

We use the JAM event generator to simulate high-energy heavy-ion collisions. Particle productions are modeled by
the excitation of hadronic resonances at low energies
and by string formation at higher energies as used in the standard microscopic transport models~\cite{RQMD1995,UrQMD1,UrQMD2,JAMorg,GiBUU,SMASH}.
JAM1~\cite{JAMorg} has been rewritten in the C++ language as JAM2.

The vector potentials for $\Lambda$ described above have been implemented
in the JAM2.1 Monte-Carlo event generator~\cite{jam2.1}.
Other updates from the version JAM2.0~\cite{JAMRQMDv} are as follows:
1) PYTHIA~8~\cite{Pythia8} library is updated to version 8.307.
2) Potentials for leading baryons are included during their formation
time with the reduced factor: 2/3 for baryons with original diquarks
and 1/3 for original quarks.
3) Collision time and ordering time has been modified following the work in Ref.~\cite{Zhao:2020yvf}.

\section{Results}
\label{sec:result}

We compute the directed flow $v_1=\langle\cos\phi\rangle$, where $\phi$ is the azimuthal angle
measured from the reaction plane, and the angle brackets indicate an average over
particles and events.
The STAR data~\cite{STAR:2017okv} in midcentral Au + Au collisions at $\sqrt{s_{NN}}=3$--$30$ GeV
from the BES program 
are compared with the result from the RQMDv mode in JAM2.1.
We chose the impact parameter $4.6<b<9.4$ fm for the midcentral Au+Au collisions.
Gaussian width $L=2.0$ fm$^2$ in Eq.~\eqref{eq:twobody} is used in the calculations.

\begin{figure}[tbhp]
\includegraphics[width=8cm]{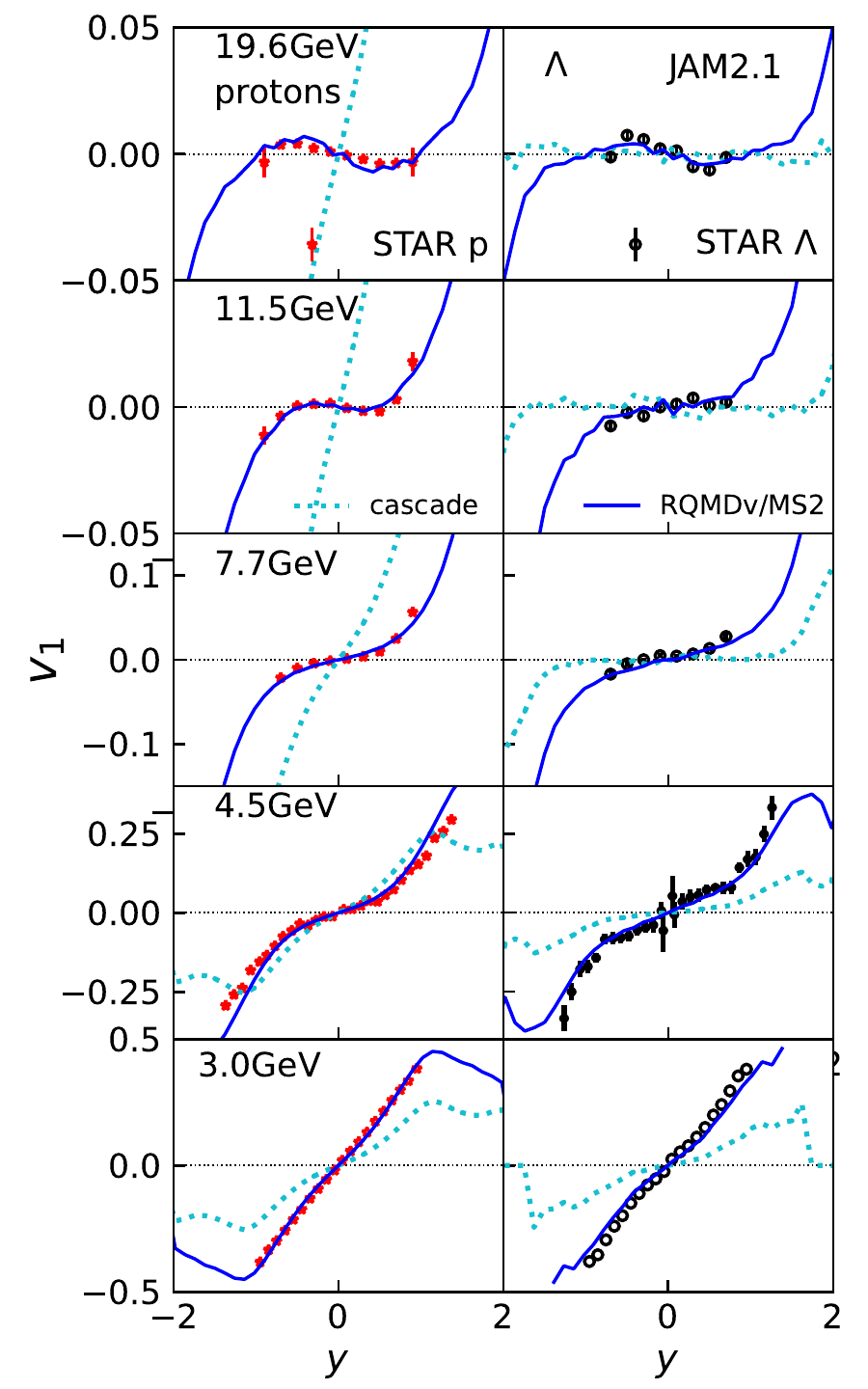}
\caption{RQMDv calculations of directed flows of protons (left panels) and $\Lambda$ (right panels) in mid-central Au + Au
collisions at $\sqrt{s_{NN}}=3.0$--$19.6$ GeV are compared with the STAR data~\cite{STAR:2017okv,STAR:2020dav,STAR:2021yiu}.
The dotted lines show the results from the Cascade mode (without potential effects),
while solid lines show the results from the RQMDv mode with the MS2 EoS\@.
}
\label{fig:v1p}
\end{figure}

First, we compare the directed flow $v_1$ of protons
and $\Lambda$ assuming that the $\Lambda$ potential is two-thirds of the nucleon
potential based on the light quark contents of nucleon and
$\Lambda$~\cite{Zhang:1999cd}.
In Fig.~\ref{fig:v1p}, we show the directed flow of protons and $\Lambda$
in midcentral Au+Au collisions at $\sqrt{s_{NN}}=3.0$--$19.6$ GeV\@.
To see the effects of potentials, we plot the cascade model predictions by dotted lines,
in which only collision terms are included.
It is clearly seen that the cascade model underestimates the proton directed flow at lower beam energies $\sqrt{s_{NN}}<11.5$ GeV
and overestimates it at higher beam energies $\sqrt{s_{NN}}>11.5$ GeV\@.
In contrast, the $\Lambda$ flow is always underestimated in the cascade model
indicating the lack of pressure arising only from the two-body collisions.
Inclusion of the potential interaction significantly improves the description of the directed flow for both protons and $\Lambda$s, and a good agreement with the STAR data~\cite{STAR:2017okv} is obtained,
where we use the soft momentum-dependent potential (MS2) from Ref.~\cite{JAMRQMDv}.
As demonstrated in Ref.~\cite{JAMRQMDv},
the transition from a negative to a positive slope of the proton directed flow
is understood by an interplay between the positive flow generated during the compression stages
and the negative flow generated dominantly in the expansion stages due to the tilted expansion.
When collision energy is low, positive flow wins because of a longer compression time,
while with increasing collision energy compression stage becomes shorter, and the expansion stage
becomes longer, which results in the net negative flow.  The turning point from a positive to negative slope
strongly depends on the strength of the interaction.

In the standard hadronic transport model,
leading hadrons within a formation time can scatter under reduced cross sections to account for
the correct Glauber-type multiple scattering~\cite{RQMD1995,UrQMD1}.
We apply a similar idea to the potential~\cite{Li:2007yd}:
we introduce potentials for the leading baryons within a formation time with reduced strength.
The leading-baryon potential predicts a more positive directed flow during the compression stages 
at $\sqrt{s_{NN}}>5$ GeV, where the string formation is dominant over the resonance
production for the particle productions.
The potential of the leading baryons improves the description of the directed flow
around $\sqrt{s_{NN}}=7.7$ GeV compared to the previous study because the string formation is dominated,
and the colliding energy $\sqrt{s_{NN}}=7.7$ GeV is still in the baryon stopping region.

The STAR data show that the turning point of the proton directed flow, as well as its shape, are the same as that of $\Lambda$ flow,
which is reproduced in the calculations.
The identical nature of the directed flow of baryons including strangeness is a remarkable finding.
If $\Xi$ directed flow may reveal to be similar to the proton directed flow,
it may be evidence of the creation of a deconfined state.  We note that a multiphase transport (AMPT) model predicts that
the $\Xi$ directed flow at mid-rapidity is the same as the $\Lambda$ flow~\cite{Nayak:2019vtn}.
A prediction for the $\Xi$ directed flow within the RQMDv approach will be reported elsewhere.

\begin{figure}[tbhp]
\includegraphics[width=8cm]{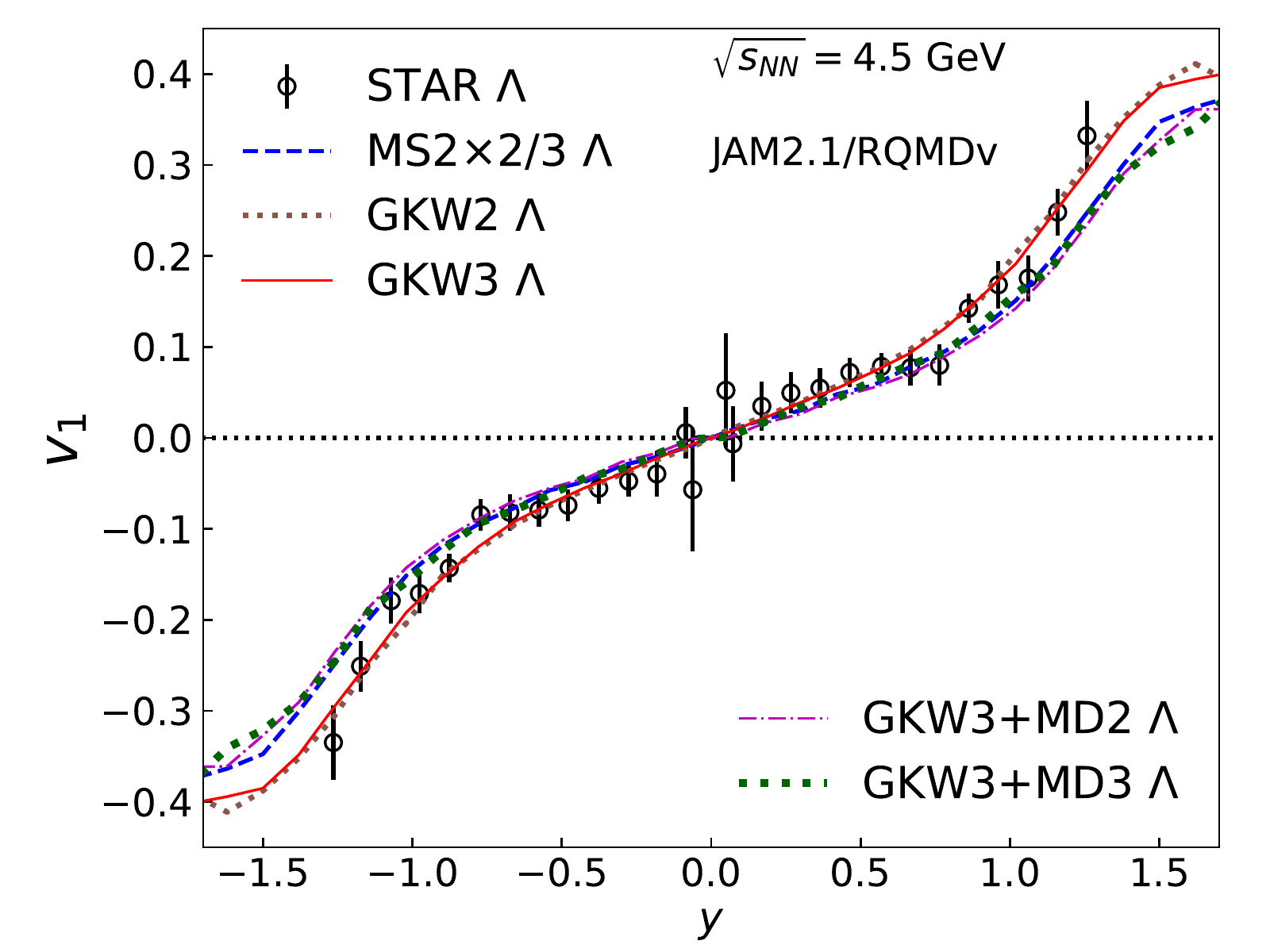}
\caption{Rapidity dependence of the directed flows of $\Lambda$ in mid-central Au + Au
collisions at $\sqrt{s_{NN}}=4.5$ GeV is compared with the STAR data~\cite{STAR:2020dav}.
The dotted lines show the results from the potential GKW2,
while the solid lines show the results from GKW3.
The dashed line corresponds to the results of the MS2 nucleon potential with the factor of 2/3,
which shows similar results from the GKW3+MD1 potential (dash-dotted line).
The GKW3+MD2 result is shown in the bold-dotted line.
}
\label{fig:v1labda45}
\end{figure}

In Fig.~\ref{fig:v1labda45}, we compare the $\Lambda$ directed flow at 4.5 GeV
for different $\Lambda$ potentials parameterized in the previous section. First, we observe that momentum-independent potentials, both GKW2 (dotted line) and GKW3 (solid line), agree with the experimental data.
Thus, the $\Lambda$ directed flow is not very sensitive to the density dependence of the potential within the density dependence given in Ref.~\cite{GKW2020}.
We omitted the results of GKW2+MD1
because it gives give almost the same results as GKW3+MD2.
In contrast, $\Lambda$ directed flow is sensitive
to the momentum dependence of the potential. 
A strongly momentum-dependent potential
GKW3 + MD2 shows a smaller $\Lambda$ directed flow,
while a weakly momentum-dependent potential GKW + MD3
predicts the same directed flow as the one in which 2/3 of nucleon potential is assumed for $\Lambda$.
We have checked that these features hold at $\sqrt{s_{NN}}=11.5$ GeV\@.
We note that the AMPT hadron cascade (AMPT-HC) model can reproduce the $\Lambda$ directed flow~\cite{Yong:2021npa}
with the momentum-independent Skyrme potential. 
Currently, we assume that all hyperons feel the same potential.
As a future work for a detailed study,
it would be important to use different hyperon potentials~\cite{Zhang:2021ddb}
for discussing the $\Sigma$ or $\Xi$ directed flows.

\begin{figure}[tbhp]
\includegraphics[width=8cm]{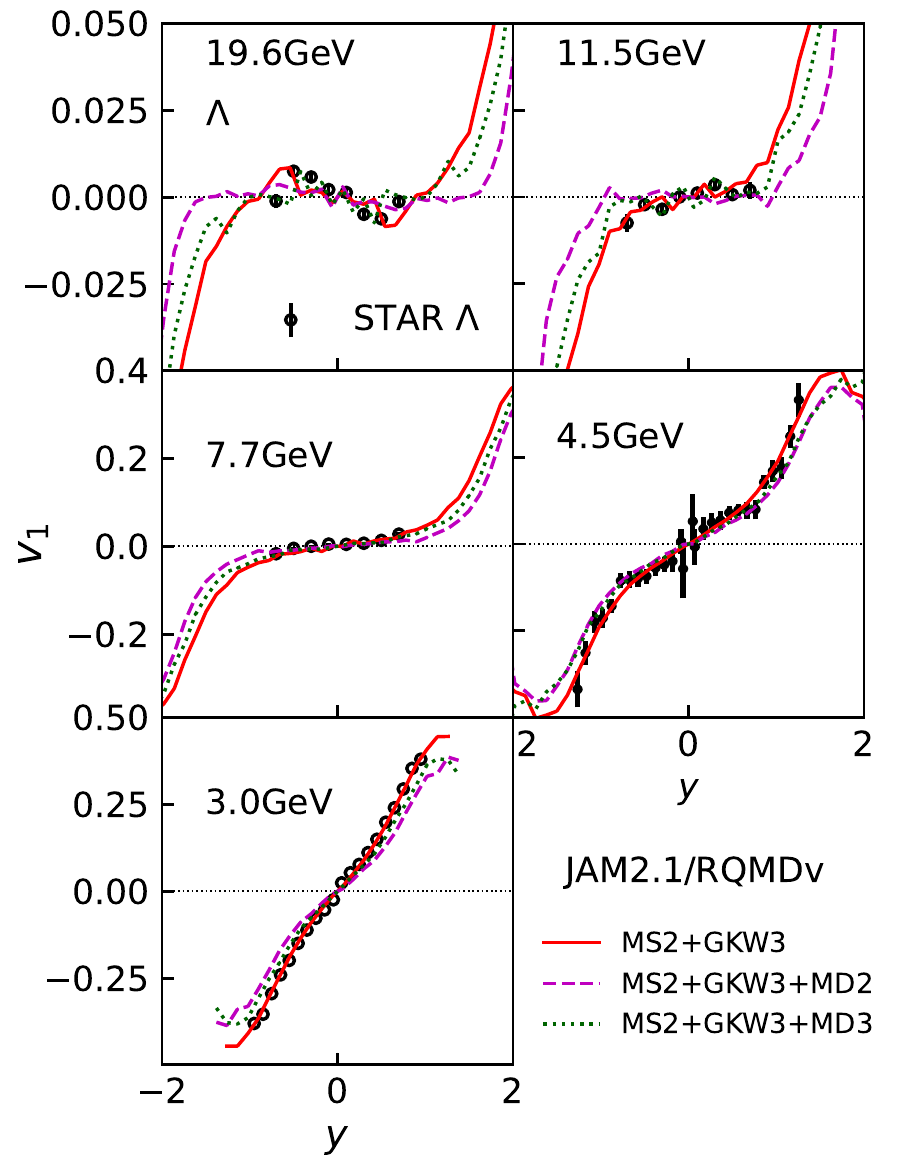}
\caption{Rapidity dependence of the  directed flow of $\Lambda$ in midcentral Au + Au collisions at
$\sqrt{s_{NN}}=3.0$--$19.6$ GeV from RQMDv are compared with the STAR data~\cite{STAR:2017okv,STAR:2020dav,STAR:2021yiu}.
The solid, dashed, and dotted lines
presents the result from MS2+GKW3, MS2+GKW3+MD1, and MS2+GWK3+MD2 EoS,
respectively.}
\label{fig:v1lambda}
\end{figure}

In Fig.~\ref{fig:v1lambda}, the directed flows of $\Lambda$
in mid-central Au + Au collisions at $\sqrt{s_{NN}}=3.0$--$19.6$ GeV
are compared with the STAR data
for different scenarios of the $\Lambda$ potentials.
The directed flow from the momentum-independent potential for $\Lambda$ (GWK3) 
and the weakly momentum-dependent potential (MD3)
show large $v_1$, especially at backward and forward rapidities,
while the strongly momentum-dependent potential (MD2) shows small $v_1$.
Thus, the momentum dependence of the $\Lambda$ potential is sensitive to the
directed flow for a wide range of incident energy.
Our calculations support the weak momentum dependence for the $\Lambda$ potential.
The momentum dependence of the 
$\Lambda$ potential deserves further investigation.
A measurement of the $\Lambda$ directed flow in the large rapidity region at higher beam energies $\sqrt{s_{NN}} >5$ GeV
further offers constraints on the momentum dependence of the $\Lambda$ potentials.

\begin{figure}[tbhp]
\includegraphics[width=8cm]{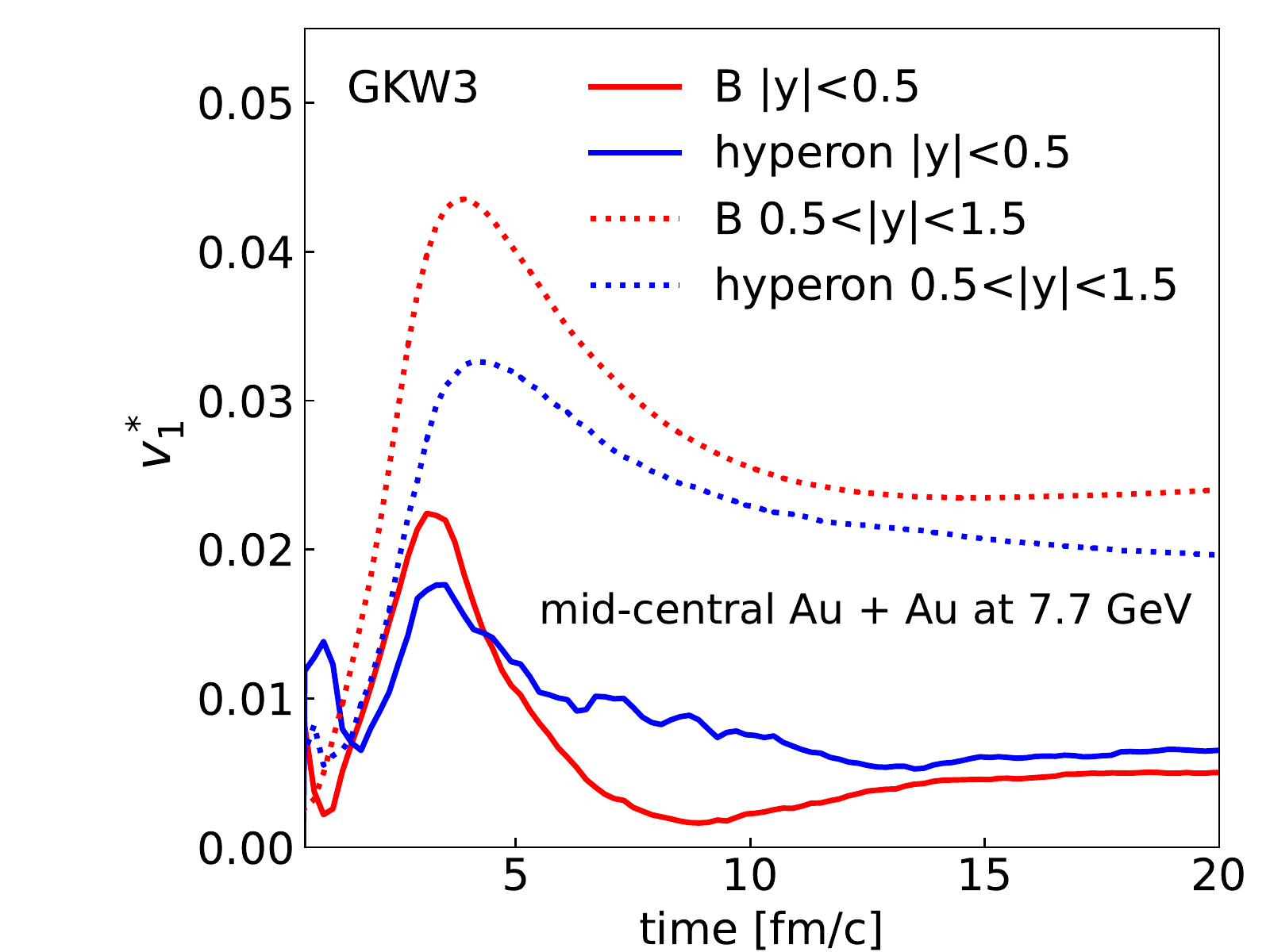}
\includegraphics[width=8cm]{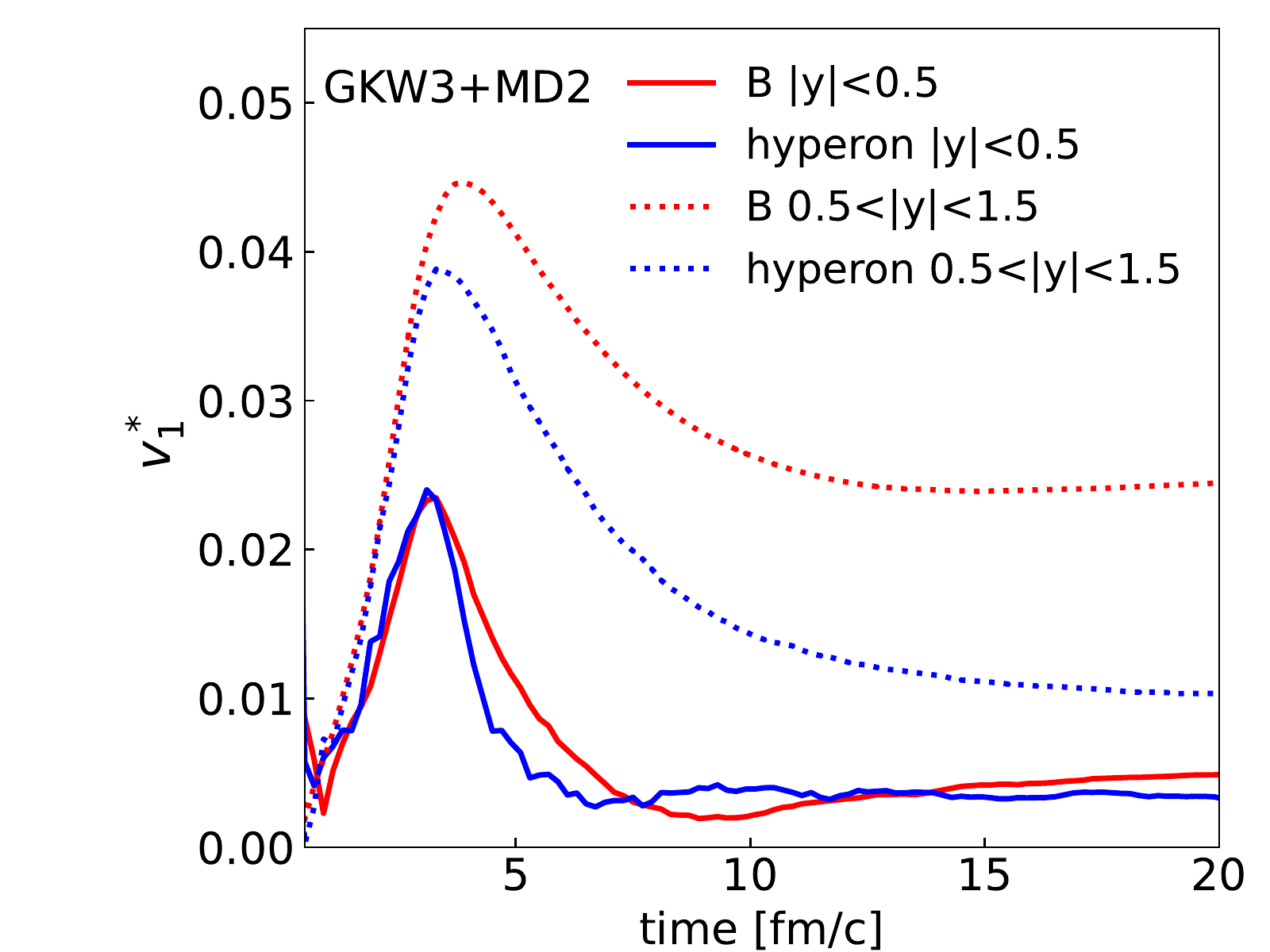}
\caption{Time evolution of the sign-weighted integrated directed flow $v^*_1$
 for nonstrange baryons (red) and hyperons (blue)
 at midrapidity $|y|<0.5$ (dotted) and forwardrapidity $0.5<|y|<1.5$.
in midcentral Au + Au collision at 7.7 GeV\@.
The upper panel shows the results from the MS2+GKW3 EoS, while the lower panel
is obtained by using the MS2+GKW3+MD2 EoS.
}
\label{fig:timeevolv1}
\end{figure}

To understand the collision dynamics for the $\Lambda$ directed flow,
we plot in Fig.~\ref{fig:timeevolv1}, the time evolution of the sign-weighted integrated directed flow
\begin{equation}
 v_1^* = \int^{y_\mathrm{max}}_{y_\mathrm{min}} dy\, v_1(y)\mathrm{sign}(y)
\end{equation}
of nonstrange baryons and hyperons for both midrapidity $|y|<0.5$ and forward-backward rapidity $0.5<|y|<1.5$
in mid-central Au + Au collisions at 7.7 GeV\@.
The upper and lower panels show results from the GKW3 and GKW3+MD2 potentials, respectively.
At midrapidity, both nonstrange baryons and hyperons show a similar behavior;
a positive directed flow is generated during the compression stages and then decreases 
due to the negative directed flow in the tilted expansion stages.
In contrast, hyperon directed flow at forward-backward rapidity show a different behavior than
the non-strange baryons: the directed flow of forward hyperons decreases more than the nonstrange baryons.
This is because most of the hyperons are produced in the hot region of matter,
while many nucleons exist close to the spectator region where the effect of tilted expansion
is weaker than the hot region.

\begin{figure}[tbhp]
\includegraphics[width=8cm]{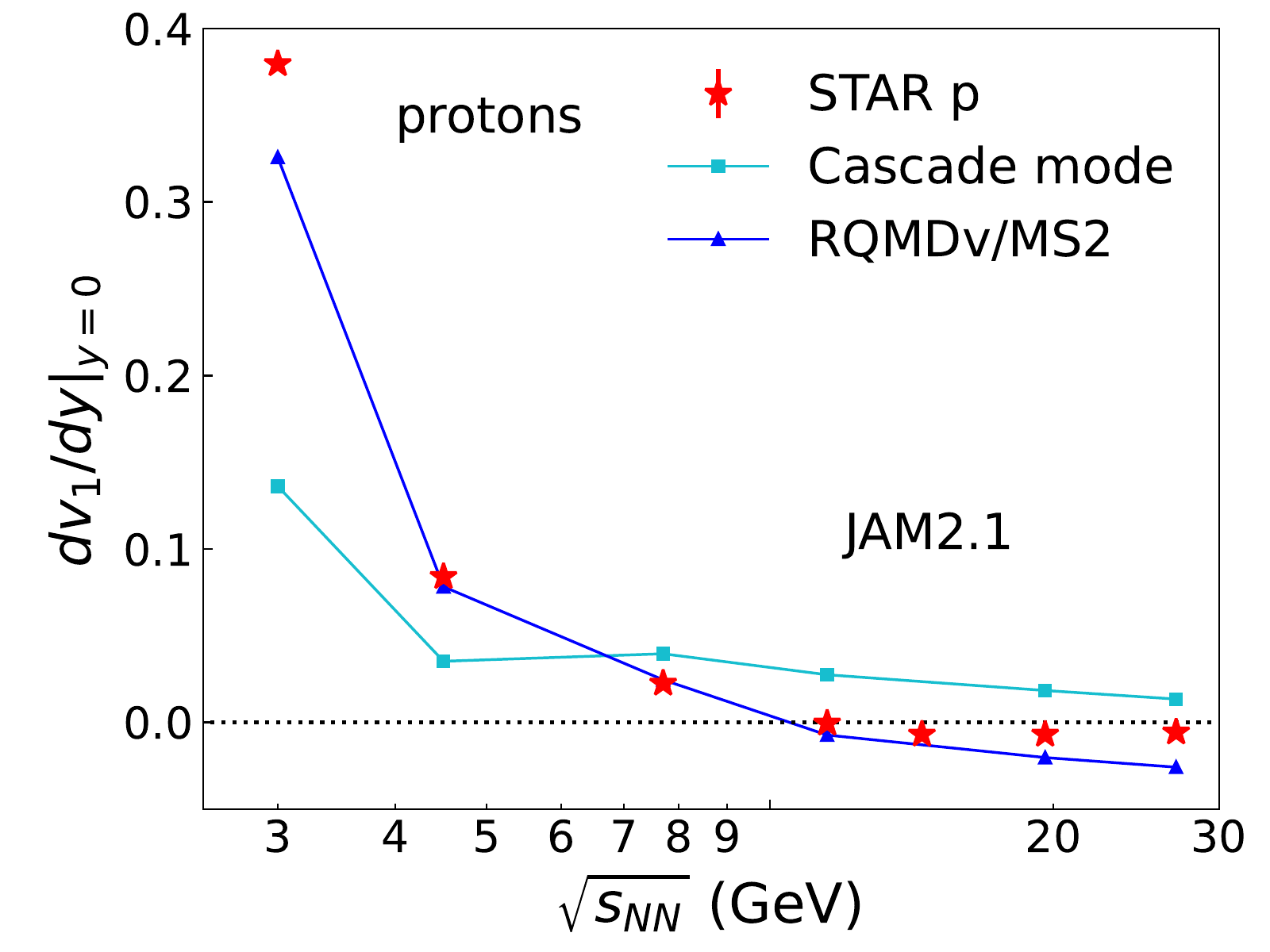}
\includegraphics[width=8cm]{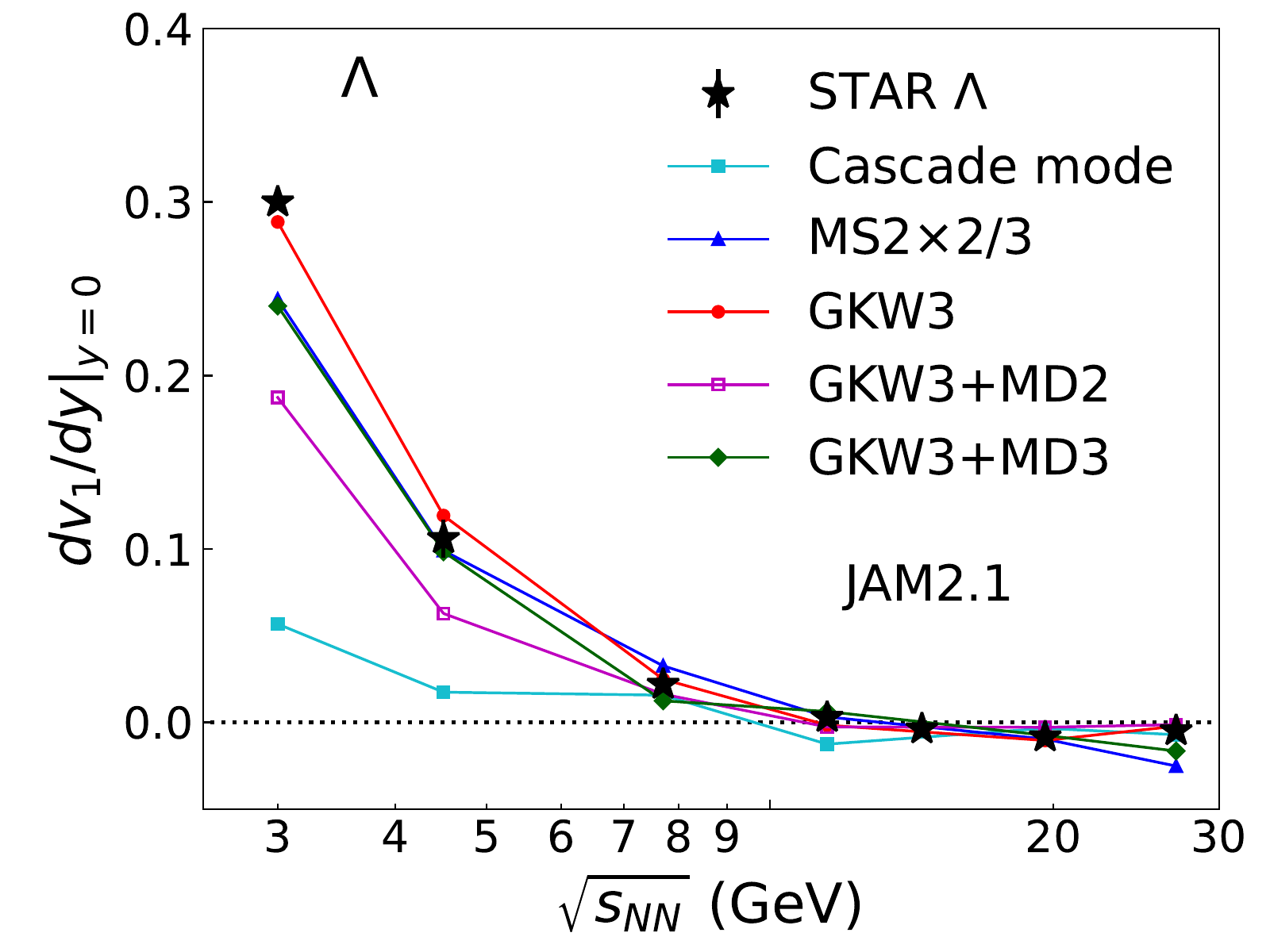}
\caption{The slopes of directed flow at mid-rapidity for protons (upper panel) and $\Lambda$ (lower panel)
in mid-central Au + Au collisions at $\sqrt{s_{NN}}=3.0-27.0$ GeV are compared with the STAR data%
~\cite{STAR:2017okv,STAR:2020dav,STAR:2021yiu}.
Cascade mode and RQMDv with MS2 EoS results are expressed by squares and triangles, respectively.
GKW3, GKW3+MD1, and GKW2+MD2 results for the slope of directed flow for  $\Lambda$ are represented by
circles, open squares, and diamonds, respectively.
}
\label{fig:v1slope}
\end{figure}

Figure~\ref{fig:v1slope} presents the beam-energy dependence of the slope of the directed flow $dv_1/dy$
at midrapidity for protons (upper panel) and $\Lambda$s (lower panel)
together with the STAR data~\cite{STAR:2017okv,STAR:2020dav,STAR:2021yiu}.
The slope $F$ is obtained by fitting the rapidity $y$ dependence
of the directed flow by $v_1(y)=Fy + F_3y^3$ over the region $|y|<0.8$.
Cascade mode (no potential) predicts a positive slope for protons at $\sqrt{s_{NN}}<30$ GeV
as is consistent with other transport model calculations~\cite{Konchakovski:2014gda}.
The inclusion of the potential interaction significantly generates a strong positive slope
at lower beam energies, while it generates a large negative slope at higher beam energies.
The slope of the $\Lambda$ directed flow becomes negative even for Cascade mode at $\sqrt{s_{NN}}>10$ GeV\@.
This is because $\Lambda$s scatter less than nucleons due to the smaller cross sections.
We see that the potential effect on the $\Lambda$ flow is significant for lower beam energies.
The beam-energy dependence of the slope of the $\Lambda$ flow is well described by the RQMDv model
with the momentum-independent GKW potential or weakly momentum-dependent
potential MD3 at $\sqrt{s_{NN}}=3.0$--$27.0$ GeV\@.
RQMDv with momentum-dependent potential predicts less slope at 3.0 GeV\@.
As it has been demonstrated in Ref.~\cite{Hartnack:1997ez},
the Gaussian width controls the interaction range of potentials in the QMD approach, where smaller width enhances the sideward flow 
$\langle p_x \rangle$. We have checked that the directed flow is enhanced with the width of $L=1.0$ fm$^2$.
However, this value does not reproduce the directed flow at the other beam 
energies unless other parameters are not tuned at the same time.
It is left for future study to understand the directed flow of $\Lambda$ at 3 GeV\@.

\section{Rapidity dependence of flow
in non-boost-invariant blast-wave model}
\label{sec:bwm}

The blast-wave model has been used to analyze the effect of transverse expansion on the observables such as
the transverse momentum spectra and the elliptic flow. 
In this section, we investigate
the implication of the similarity of the $\Lambda$ flow with the proton flow discovered by the STAR collaboration
within a hydrodynamic scenario by using the blast-wave model.
The blast-wave analysis provides
a complementary study to the nonequilibrium microscopic transport approach RQMDv.
It should be emphasized that the blast-wave model
provides a simple fitting of spectra and anisotropic flows, but does not provide detailed information about the dynamical aspects of nuclear collisions, while the microscopic transport model of the RQMD model does.

\subsection{Blast-wave model}

The non-boost-invariant formula for the blast-wave model
can be found in Ref.~\cite{Dobler:1999ju}.
It is shown that both transverse momentum and rapidity distributions
in Au + Au collisions at the BNL Alternating Gradient Synchrotron energies and Pb + Pb collisions at the CERN Super Proton Synchrotron energies are well
fitted by the model~\cite{Rode:2018hlj}.
Applications to the elliptic flow at midrapidity
can be found in Refs.~\cite{Huovinen:2001cy,STAR:2001ksn,Sun:2014rda}.
A simple formula for the directed flow was proposed in Ref.~\cite{Voloshin:1996nv}.
We apply the non-boost-invariant blast-wave model of
Ref.~\cite{Dobler:1999ju} to compute the rapidity dependence of the anisotropic
flows, which may be obtained as
\begin{equation}
v_n(y)=\frac{V_n(y)}{V_0(y)}
\end{equation}
with
\begin{align}
V_n(y)&=
  \int^{2\pi}_0d\phi \cos(n\phi)
\int^{\eta_\mathrm{max}}_{-\eta_\mathrm{max}} d\eta
 \int_0^{R(\eta)} rdr \nonumber\\
& \times \, \int m_\perp dm_\perp \alpha
 \exp\left(\frac{\mu - \alpha \cosh\rho}{T}\right)
I_n\left(\beta\right) \,,
\end{align}
where 
$\alpha=m_\perp\cosh(y-\eta)$,
$\beta=p_\perp\sinh(\rho)/T$,
$\eta = \tanh^{-1}(z/t)$ is the space-time rapidity,
and 
$\rho=\tanh^{-1}v_\perp$ is related to the collective transverse fluid velocity $v_\perp$.
The chemical potential for the baryon with the baryon number $B$ and the strangeness $S$
is fixed by $\mu=B\mu_B+S\mu_S$.
The $\eta$-dependence of the radius is taken to be
\begin{equation}
R(\eta) = R_0\sqrt{1-\frac{\eta^2}{\eta_\mathrm{max}^2}}.
\end{equation}
We make an ansatz for the shape of the directed flow neglecting
the higher-order flows
\footnote{For example, the second-order flow can be added $\rho_2(\eta)\cos2\phi$ with $\rho_2(\eta)=a_2 + b_2\eta^2 + c_2\eta^4$
to investigate the rapidity dependence of the elliptic flow.}
\begin{equation}
\rho(\eta,\phi,r) = \rho_0(\eta,r)[1 + \rho_1(\eta)\cos\phi], 
\end{equation}
where $\rho_0$ controls the transverse radial flow profile,
and its dependence on the radius $r$ and
space-time rapidity $\eta$ is assumed to be
\begin{equation}
\rho_0(\eta,r)  =
\rho_0\left(\frac{r}{R_0}\right)\sqrt{1-\frac{\eta^2}{\eta_\mathrm{max}^2}}.
\end{equation}
We assume the following shape
for the directed flow coefficient:
\begin{equation}
\rho_1(\eta) =  a \eta + b \eta^3 + c\eta^5.
\end{equation}

\begin{figure}[htbp]
\includegraphics[width=8cm]{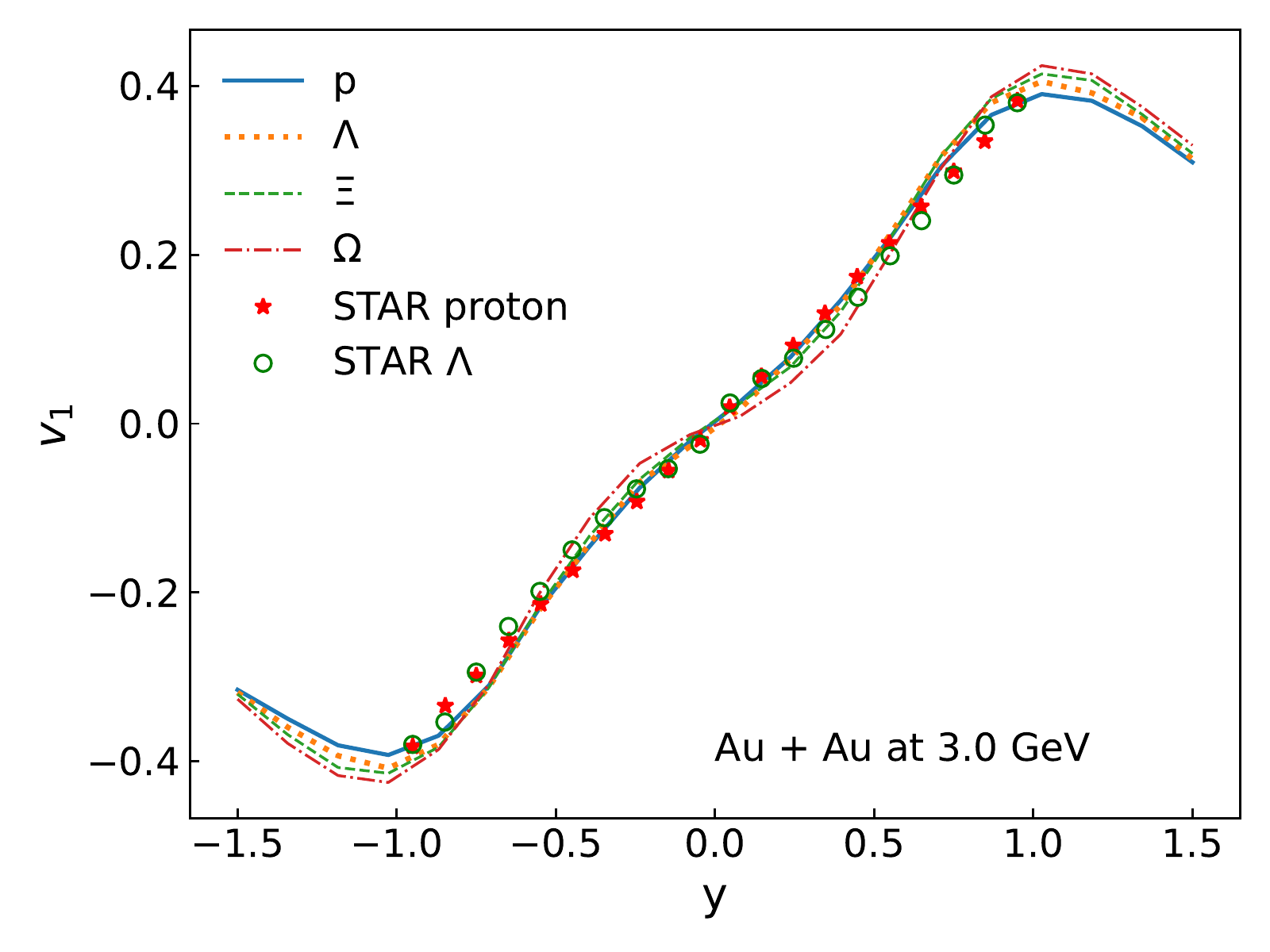}
\includegraphics[width=8cm]{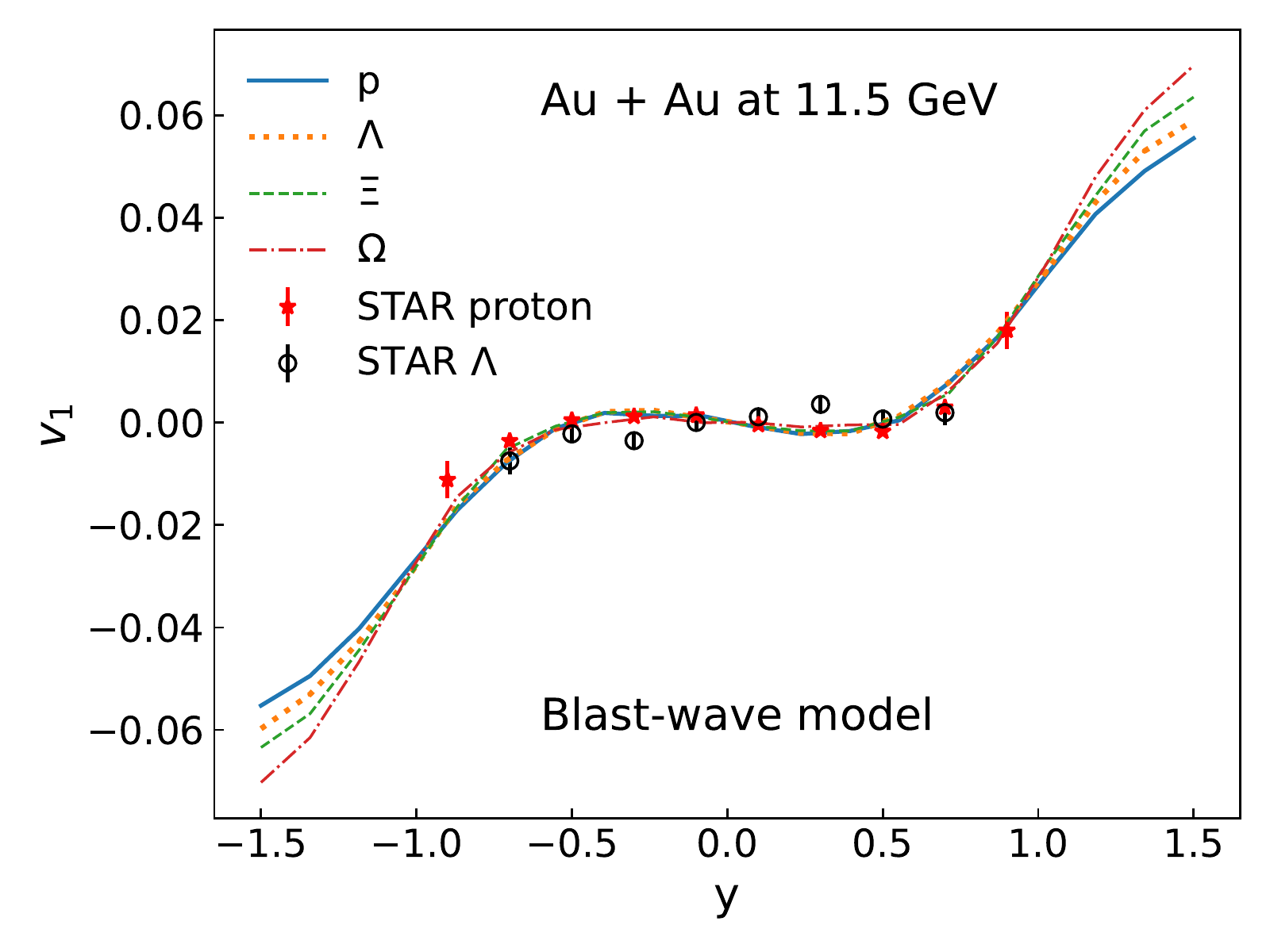}
\caption{Blast-wave model fit of the rapidity dependence of the directed flows of protons (solid line)
 and hyperons, $\Lambda$ (dotted), $\Xi$ (dashed), and $\Omega$ (dash-dotted)
 in midcentral Au + Au collisions at $\sqrt{s_{NN}}=3.0$ and 11.5 GeV
to the STAR data~\cite{STAR:2021yiu,STAR:2017okv}.
}
\label{fig:bwev1}
\end{figure}

Figure~\ref{fig:bwev1} shows the blast-wave fits of the rapidity dependence of the directed flows
for protons, $\Lambda$s, $\Xi$s, and $\Omega$s
together with the STAR data for 3 GeV (upper panel) and 11.5 GeV (lower panel).
We here considered common parameters of the profile and the flow for different hadrons
assuming that the hadrons fully thermalize
to have a common freeze-out temperature and obey a single collective flow.
For more realistic calculations, we could use different kinetic freeze-out temperatures
for the multistrange hadrons because they are likely to decouple
from the system earlier than non-strange baryons%
~\cite{vanHecke:1998yu,Takeuchi:2015ana}.

We fixed the parameters as
  $\eta_\mathrm{max}=1.1$,
  $\mu_B=0.7$ GeV, $\mu_S=0.06$ GeV,
  $T=0.08$ GeV,
  $\rho_0=0.7$,
  $a =0.8$, $b= 1.0$, and $c=5.0$ for 3 GeV, 
 and
  $\eta_\mathrm{max}=1.645$,
  $T=0.12$ GeV,
  $\rho_0=0.8$,
  $\mu_B=0.2$ GeV,  $\mu_S=0.06$ GeV,
  $a =-0.05$, $b= 0.05$, and $c=0.1$ for 11.5 GeV\@.
It is seen that the blast-wave model predicts that the hyperon directed flows show similar rapidity dependence except at a very large rapidity.
We see that the rapidity dependence of the $\Omega$ baryon reveals the mass effect slightly.
For a more detailed analysis, we need to include the contributions from hadron resonances, which is left for future work.

\subsection{Quark recombination}
\label{sec:reco}

We now consider the directed flow in the quark recombination model within the blast-wave model.
Within the $\delta$-function approximation (the limit of the zero-momentum spread of quark momentum fractions)
for the hadron wave function, and
neglecting higher order anisotropic flows, 
baryon directed flows from quark recombination
are given by~\cite{Lin:2003jy,Molnar:2003ff,Fries:2003kq}
\begin{equation}
 v_1^B(y) = \frac{v_1^a + v_1^b+ v_1^c + 3v_1^a v_1^b v_1^c}
  {1 + 2 v_1^a v_1^b
     + 2 v_1^b v_1^c
     + 2 v_1^c v_1^a},
\end{equation}
where $v_1^{a,b,c}$ are the directed flows of
valence quarks.
If $v_1^a = v_1^b=v_1^c \;(\equiv v_1^q)$ and $v_1^q \ll 1$, we have a simple quark scaling of the flow
\begin{equation}
v_1^B(y)= 3 v_1^q,
\end{equation}
where the quark flow $v_1^q$ is calculated using the quark momentum
being one-third of the baryon momentum: $\bm{p}_q = \bm{p}_B/3$.

\begin{figure}[tbhp]
\includegraphics[width=8cm]{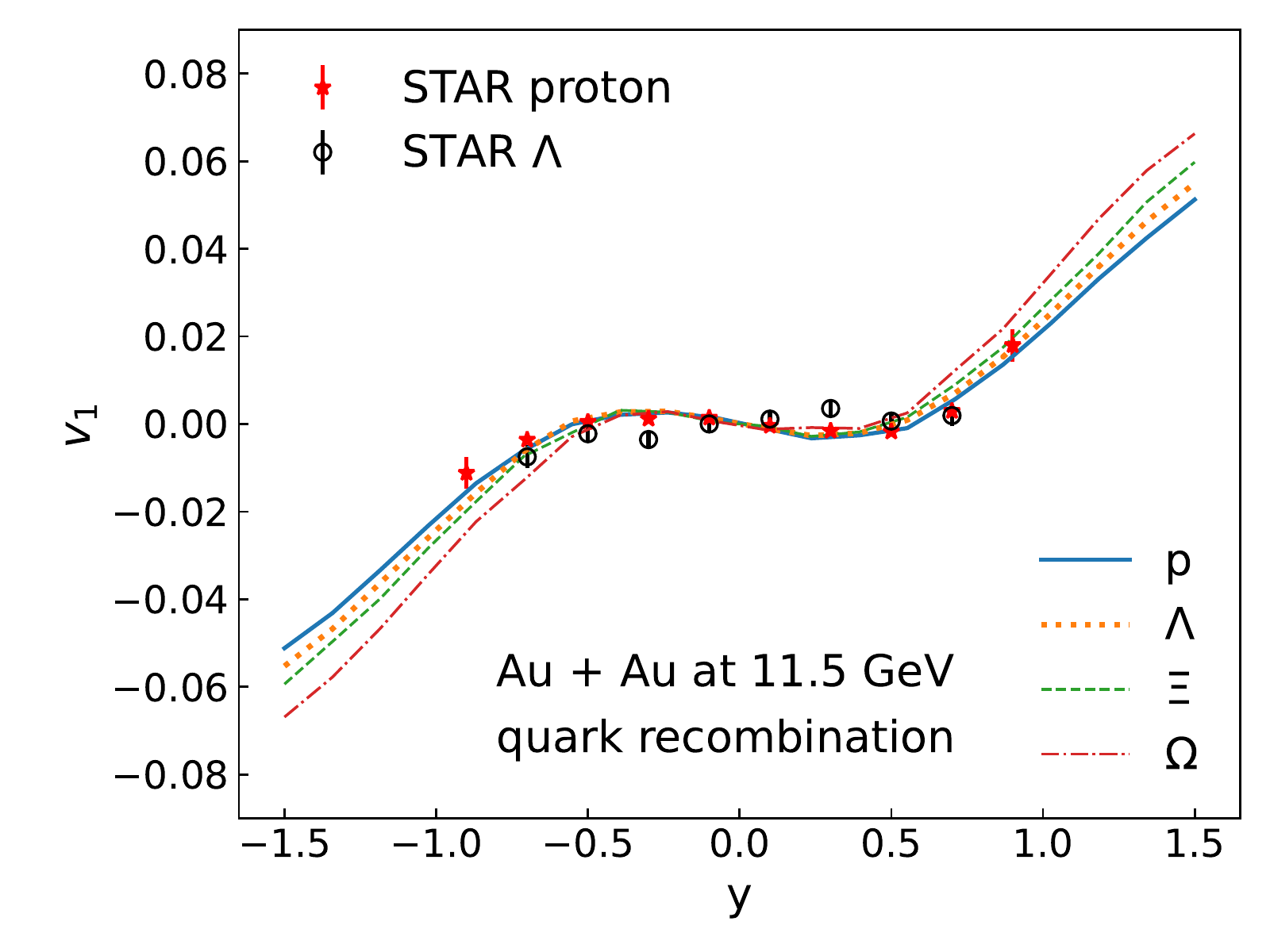}
\caption{Quark recombination model fit of the rapidity dependence of the directed flow
for protons and hyperons.
The parameters of $a=-0.13$, $b= 0.06$, and $c=0.12$ are used.
The STAR data are taken from~\cite{STAR:2017okv}.
}
\label{fig:recov1}
\end{figure}

In Fig.~\ref{fig:recov1}, we compare the rapidity dependence of the directed flows
of protons and various hyperons within the quark recombination model.
We assumed the constituent quark masses $m_{u} = 0.336~\GeV$,
$m_d=0.34~\GeV$, and $m_s = 0.486~\GeV$.
The quark recombination predictions are consistent with the STAR data for protons and $\Lambda$s at $|y|<1.0$.
However, it predicts different rapidity dependence
for different hyperons.
The mass effects on the rapidity dependence of the flow come from
the longitudinal momentum of quarks, which is computed by one-third of the baryon momentum: $p_{q,z}=(m_{\perp}\sinh y)/3$,
where $m_\perp=\sqrt{m_h^2+p_{h,\perp}^2}$ denotes the transverse mass of the hadron.
Thus, for larger hadron mass $m_h$, we pick larger rapidity
of quarks, which makes the different rapidity dependence of the directed flow of hyperons. 
The data for $\Xi$ and $\Omega$ at larger rapidities 
may be useful to distinguish the quark coalescence
from other pictures by seeing the rapidity dependence of quark degrees of freedom.

\section{Summary}
\label{sec:summary}

We have investigated the rapidity and beam-energy dependence of the $\Lambda$ directed flow
from heavy-ion collisions at $\sqrt{s_{NN}}=3.0$--$30.0$ GeV
within a non-equilibrium microscopic transport model JAM/RQMDv
with different assumptions of the $\Lambda$ potentials.
The RQMDv results agree with the STAR data on the beam-energy and rapidity dependence of the $\Lambda$ directed flow.
We compared the density- and momentum-dependent $\Lambda$ potentials calculated from $\chi$EFT with
two- and three-body interactions, which suppresses the appearance of $\Lambda$ hyperons in neutron stars. 
It is found that three-body interactions do reproduce
the $\Lambda$ directed flow for a wide range of beam energies
as well as the rapidity dependence.
This is the first examination of the strong repulsion of
the $\Lambda$ potential in nuclear matter from the heavy-ion data.
However, at the same time, we found that the directed flow of $\Lambda$ can also be reproduced by the $\Lambda$ potential only with the two-body interactions having a weaker repulsion at high densities.
On the other hand, the $\Lambda$ directed flow is strongly affected by the momentum dependence, especially in a large rapidity region.
Most of $\Lambda$s are produced in a dense part of the matter and
more susceptible to a tilted expansion of matter than the nucleons.
Thus, the STAR BES data do not strongly constrain the density dependence of the $\Lambda$ potential by the directed flow, but the momentum dependence of the hyperon
potential may be fixed by the directed flow data.

As a complementary study,
we analyze the directed flow by the blast-wave model,
which provides an insight into spectra from a hydrodynamic picture.
The blast-wave model predicts similar rapidity dependence of the 
directed flow for protons, $\Lambda$s, and $\Xi$s,
while $\Omega$ directed flow shows a slight deviation
from other hyperons due to a heavier mass.
On the other hand, quark coalescence predicts large mass
effects at forward-backward rapidities in the rapidity dependence of the hyperon directed flows
because of kinematical effects: the longitudinal momenta of quarks become large for heavy baryons.

Systematic studies of directed as well as elliptic flows of hyperons, $\Lambda$, $\Xi$, and $\Omega$,
may provide further constraints for the EoS of highly dense matter including strangeness.

\begin{acknowledgments}
The authors would like to thank Prof. Hideo Suganuma and Dr. Tokuro Fukui for useful suggestions.
This work was supported in part by the
Grants-in-Aid for Scientific Research from JSPS
(Nos. JP21K03577,
 JP19H01898, 
and JP21H00121
).
\end{acknowledgments}

\appendix

\section{Momentum-dependent vector potential in the nuclear matter}
\label{app:potential}

We consider the vector-type density-dependent potential $U^\mu_\sk$ and momentum-dependent potential
$U_m^\mu$ in the nuclear matter.
For the momentum-dependent vector-potential implementation,
the energy density of the system
has the following form~\cite{RBUUp}:
\begin{equation}
\epsilon = \int d^3p \left( e^* + U_m^0 -
 \frac{1}{2}\frac{p^*_{\mu}}{e^*}U^\mu_m(p)\right) f(p)
  + \int_0^{\rho} U^0_\sk(\rho')d\rho' \,,
\end{equation}
where kinetic energy and momentum are defined as $e^*=\sqrt{m_N^{2}+\bm{p}^{*2}}$
 and $p^{*^\mu}=p^\mu - U^\mu = p^\mu - U_\sk^\mu - U_m^\mu$.
Let us consider the terms
\begin{equation}
 \int \frac{d^3p}{e^{*}}
\bm{p}^*
\cdot \bm{U}_m 
= \int \frac{d^3p}{e^{*}}\frac{d^3p'}{e^{*'}}
    \bm{p}^*\cdot\bm{p}^{*'}
     \frac{f(x,p')}{1+[(\bm{p}-\bm{p}')/\mu]^2}
\end{equation}
and 
\begin{equation}
e^*=\sqrt{m_N^{2}+(\bm{p}-\bm{U}_\sk(\rho)-\bm{U}_m(p))^2}.
\end{equation}
Because of the rotational invariance in the nuclear matter, 
the spatial part of the density-dependent potential is zero,
$\bm{U}_\sk=0$, and
the momentum-dependent part of the vector potential must
have the following form:
\begin{equation}
 U_m^\mu = (U^0_m,\, \bm{U}_m)
=\left(U^0_m,\,\frac{\bm{p}}{p}U_m(p)\right)\,,
\end{equation}
where $p=|\bm{p}|$, and thus $\bm{p}^*=\bm{p}(1-U_m(p)/p)$.
The energy density for the nuclear matter at zero temperature is given by
\begin{equation}
 \epsilon = \int_0^{p_f} d^3p \left( e^* + \frac{1}{2}U_m^0
 -\frac{1}{2}(p-U_m)U_m \right)
  + \int_0^{\rho} U^0_\sk(\rho')d\rho' \,,
\end{equation}
where $e^* = \sqrt{m^{*2}+(p-U_m)^2}$.
After performing the angular integral,
\begin{equation}
 U_m(p)=2\pi \frac{\mu^2}{p}\int^{p_f}_0 dp'p'\frac{p^{*'}}{e^{*'}}\left(
   \frac{A}{4pp'}\ln\left|\frac{A+2pp'}{A-2pp'} \right| -1
\right),
\end{equation}
where $A=p^2+p^{'2}+\mu^2$.
The optical potential is defined as the difference between the single-particle energy
and the kinetic energy:
\begin{equation}
\label{eq:VecOptPot}
U_\mathrm{opt}(\rho,p) = p^0 - \sqrt{m_N^2 + \bm{p}^2}
                  =e^* + U_\sk^0 + U_m^0 - \sqrt{m_N^2 + \bm{p}^2}\,.
\end{equation}
The non-relativistic limit is obtained by taking $\bm{U}_m(p)=0$.
To avoid another integral,
the pressure at zero temperature  can be calculated by using the energy density
and single-particle energy $p^0$:
\begin{equation}
 P = \rho^2 \frac{\partial}{\partial\rho}\left( \frac{\epsilon}{\rho}\right)
  =\rho \frac{\partial \epsilon}{\partial\rho} - \epsilon
   = \rho\, p^0(p_f) - \epsilon
\end{equation}
The incompressibility is given by
\begin{equation}
K=9\rho\frac{\partial p^0}{\partial\rho}=
 9\rho\left[
  \frac{p_f^*}{e_f^*}
  \left(
    \frac{p_f}{3\rho}
   -\frac{\partial U_m}{\partial \rho}\right)
  + \frac{\partial U_\sk^0}{\partial\rho}
  + \frac{\partial U_m^0}{\partial \rho}
 \right].
\end{equation}
We obtained $K$ numerically by using a finite difference with the fourth-order accuracy.


\begin{thebibliography}{99}

\bibitem{Danielewicz:2002pu}
  P.~Danielewicz, R.~Lacey and W.~G.~Lynch,
  Science {\bf 298}, 1592 (2002).

\bibitem{Oertel:2016bki}
M.~Oertel, M.~Hempel, T.~Kl\"ahn and S.~Typel,
Rev. Mod. Phys. \textbf{89}, no.1, 015007 (2017).


\bibitem{Poskanzer:1998yz}
A.~M.~Poskanzer and S.~A.~Voloshin,
Phys. Rev. C \textbf{58}, 1671-1678 (1998).

\bibitem{Rischke:1995pe}
  D.~H.~Rischke, Y.~Pursun, J.~A.~Maruhn, H.~Stoecker and W.~Greiner,
  Acta Phys.\ Hung.\ A {\bf 1}, 309 (1995).

\bibitem{Brachmann:1999xt}
J.~Brachmann, S.~Soff, A.~Dumitru, H.~Stoecker, J.~A.~Maruhn, W.~Greiner,
L.~V.~Bravina and D.~H.~Rischke,
Phys.\ Rev.\ C {\bf 61}, 024909 (2000).

\bibitem{Csernai1999} L. P. Csernai and D. Rohrich, Phys. Lett. {\bf B} 458, 454 (1999).

\bibitem{Li:1998ze}
  B.~A.~Li and C.~M.~Ko,
  Phys.\ Rev.\ C {\bf 58}, 1382 (1998).

\bibitem{STAR2014}
L. Adamczyk et al. [STAR], Phys. Rev. Lett. \textbf{112}, 162301 (2014).
\bibitem{STAR2016}
P. Shanmuganathan et al. [STAR], Nucl. Phys. A \textbf{956}, 260 (2016).


\bibitem{STAR:2017okv}
L.~Adamczyk \textit{et al.} [STAR],
Phys. Rev. Lett. \textbf{120}, no.6, 062301 (2018).

\bibitem{STAR:2020dav}
J.~Adam \textit{et al.} [STAR],
Phys. Rev. C \textbf{103}, no.3, 034908 (2021).

\bibitem{STAR:2021yiu}
M.~S.~Abdallah \textit{et al.} [STAR],
Phys. Lett. B \textbf{827}, 137003 (2022).








\bibitem{Steinheimer:2014pfa}
J.~Steinheimer, J.~Auvinen, H.~Petersen, M.~Bleicher and H.~Stoecker,
Phys.\ Rev.\ C {\bf 89}, no. 5, 054913 (2014).

\bibitem{Konchakovski:2014gda}
V.~P.~Konchakovski, W.~Cassing, Y.~B.~Ivanov and V.~D.~Toneev,
Phys.\ Rev.\ C {\bf 90}, no. 1, 014903 (2014).


\bibitem{Ivanov:2014ioa}
  Y.~B.~Ivanov and A.~A.~Soldatov,
  Phys.\ Rev.\ C {\bf 91}, no. 2, 024915 (2015);
  Y.~B.~Ivanov and A.~A.~Soldatov,
  Eur.\ Phys.\ J.\ A {\bf 52}, no. 1, 10 (2016);


\bibitem{Nara:2016hbg}
  Y.~Nara, H.~Niemi, J.~Steinheimer and H.~Stoecker,
  Phys.\ Lett.\ B {\bf 769}, 543 (2017).


\bibitem{JAMRQMDv}
Y.~Nara and A.~Ohnishi,
Phys. Rev. C \textbf{105}, no.1, 014911 (2022).


\bibitem{GKW2020}
D. Gerstung, N. Kaiser, W. Weise, Eur. Phys. J. A \textbf{56}, 175 (2020).

\bibitem{HyperonPuzzle}
P. Demorest \textit{et al.},
Nature \textbf{467}, 1081-1083 (2010);
%
E. Fonseca \textit{et al.},
Astrophys. J. \textbf{832}, 167 (2016);
%
J. Antoniadis \textit{et al.},
Science \textbf{340}, 6131 (2013);
%
H. T. Cromartie \textit{et al.} [NANOGrav],
Nature Astron. \textbf{4}, 72 (2019);
%
M. C. Miller \textit{et al.}, 
Astrophys. J. Lett. \textbf{918}, L28 (2021).
%



\bibitem{Bertsch:1988ik}
G.~F.~Bertsch and S.~Das Gupta,
Phys.\ Rept.\  {\bf 160}, 189 (1988).

\bibitem{Cassing:1990dr}
W.~Cassing, V.~Metag, U.~Mosel and K.~Niita,
Phys.\ Rept.\  {\bf 188}, 363 (1990).

\bibitem{Ko:1987gp}
C.~M.~Ko, Q.~Li and R.~C.~Wang,
Phys.\ Rev.\ Lett.\  {\bf 59}, 1084 (1987);


\bibitem{Blaettel:1993uz}
B.~Blaettel, V.~Koch and U.~Mosel,
Rept.\ Prog.\ Phys.\  {\bf 56}, 1 (1993).


\bibitem{GiBUU}
  O.~Buss {\it et al.},
  Phys.\ Rept.\  {\bf 512}, 1 (2012).

\bibitem{Aichelin:1991xy}
J.~Aichelin,
Phys.\ Rept.\  {\bf 202}, 233 (1991).

\bibitem{RQMD1989}
H.~Sorge, H.~Stoecker and W.~Greiner,
Annals Phys.\  {\bf 192}, 266 (1989).


\bibitem{UrQMD1}
  S.~A.~Bass {\it et al.},
  Prog.\ Part.\ Nucl.\ Phys.\  {\bf 41}, 255 (1998).

\bibitem{Isse:2005nk}
M.~Isse, A.~Ohnishi, N.~Otuka, P.~K.~Sahu and Y.~Nara,
Phys.\ Rev.\ C {\bf 72}, 064908 (2005).


\bibitem{Aichelin:2019tnk}
J.~Aichelin, E.~Bratkovskaya, A.~Le F\`evre, V.~Kireyeu, V.~Kolesnikov, Y.~Leifels, V.~Voronyuk and G.~Coci,
Phys. Rev. C \textbf{101}, no.4, 044905 (2020).




\bibitem{Danielewicz:1998vz}
  P.~Danielewicz, R.~A.~Lacey, P.~B.~Gossiaux, C.~Pinkenburg, P.~Chung,
J.~M.~Alexander and R.~L.~McGrath,
  Phys.\ Rev.\ Lett.\  {\bf 81}, 2438 (1998).

\bibitem{Rai:1999hz}
  G.~Rai {\it et al.} [E895 Collaboration],
  Nucl.\ Phys.\ A {\bf 661}, 162 (1999).


\bibitem{Hillmann:2018nmd}
  P.~Hillmann, J.~Steinheimer and M.~Bleicher,
  J.\ Phys.\ G {\bf 45}, no. 8, 085101 (2018).


\bibitem{Oliinychenko:2022uvy}
D.~Oliinychenko, A.~Sorensen, V.~Koch and L.~McLerran,
[arXiv:2208.11996 [nucl-th]].

\bibitem{Steinheimer:2022gqb}
J.~Steinheimer, A.~Motornenko, A.~Sorensen, Y.~Nara, V.~Koch and M.~Bleicher,
[arXiv:2208.12091 [nucl-th]].





\bibitem{TMEP:2022xjg}
H.~Wolter \textit{et al.} [TMEP],
Prog. Part. Nucl. Phys. \textbf{125}, 103962 (2022).

\bibitem{Dobler:1999ju}
H.~Dobler, J.~Sollfrank and U.~W.~Heinz,
Phys. Lett. B \textbf{457}, 353-358 (1999).

\bibitem{LambdaPot-NonRel}
S.~Balberg and A.~Gal,
Nucl. Phys. A \textbf{625}, 435 (1997);
%
D.~E.~Lanskoy and Y.~Yamamoto,
Phys. Rev. C \textbf{55}, 2330 (1997);
%
M.~Baldo, G.~F.~Burgio and H.~J.~Schulze,
Phys. Rev. C \textbf{61}, 055801 (2000).


\bibitem{LambdaPot-Rel}
N.~K.~Glendenning and S.~A.~Moszkowski,
Phys. Rev. Lett. \textbf{67}, 2414 (1991);
J.~Schaffner and I.~N.~Mishustin,
Phys. Rev. C \textbf{53}, 1416 (1996);
C.~Ishizuka, A.~Ohnishi, K.~Tsubakihara, K.~Sumiyoshi and S.~Yamada,
J. Phys. G \textbf{35}, 085201 (2008);
%
H.~Shen, H.~Toki, K.~Oyamatsu and K.~Sumiyoshi,
Astrophys. J. Suppl. \textbf{197}, 20 (2011).


\bibitem{HyperonPuzzleSolution}
S.~Nishizaki, T.~Takatsuka and Y.~Yamamoto,
Prog. Theor. Phys. \textbf{108}, 703 (2002);
%
J.~Rikovska-Stone, P.~A.~M.~Guichon, H.~H.~Matevosyan and A.~W.~Thomas,
Nucl. Phys. A \textbf{792}, 341 (2007);
%
S.~Weissenborn, D.~Chatterjee and J.~Schaffner-Bielich,
Phys. Rev. C \textbf{85}, 065802 (2012)
[erratum: Phys. Rev. C \textbf{90}, 019904 (2014)];
%
T.~Miyatsu, S.~Yamamuro and K.~Nakazato,
Astrophys. J. \textbf{777}, 4 (2013).
%
H.~Togashi, E.~Hiyama, Y.~Yamamoto and M.~Takano,
Phys. Rev. C \textbf{93}, 035808 (2016).

\bibitem{chiralEFT}
S. Weinberg,
Phys. Rev. Lett. \textbf{17}, 616 (1966);
%
E. Epelbaum, H. W. Hammer and U. G. Meissner,
Rev. Mod. Phys. \textbf{81}, 1773 (2009);
%
R. Machleidt and D. R. Entem,
Phys. Rept. \textbf{503}, 1 (2011).


\bibitem{Haidenbauer-NLO}
J.~Haidenbauer, S.~Petschauer, N.~Kaiser, U.~G.~Meissner, A.~Nogga and W.~Weise,
Nucl. Phys. A \textbf{915}, 24-58 (2013);
J.~Haidenbauer, U.~G.~Mei\ss{}ner and A.~Nogga,
Eur. Phys. J. A \textbf{56}, no.3, 91 (2020).

\bibitem{Kohno:2018gby}
M.~Kohno,
Phys. Rev. C \textbf{97}, no.3, 035206 (2018)
doi:10.1103/PhysRevC.97.035206
[arXiv:1802.05388 [nucl-th]].

\bibitem{Decuplet}
S.~Petschauer, J.~Haidenbauer, N.~Kaiser, U.~G.~Mei\ss{}ner and W.~Weise,
Nucl. Phys. A \textbf{957}, 347-378 (2017);
J.~Haidenbauer, S.~Petschauer, N.~Kaiser, U.~G.~Mei\ss{}ner and W.~Weise,
Eur. Phys. J. C \textbf{77}, no.11, 760 (2017).

\bibitem{Tews2017}
I. Tews, J. M. Lattimer, A. Ohnishi, E. E. Kolomeitsev,
Astrophys. J. \textbf{848}, 105 (2017).

\bibitem{RBUUp}
  K.~Weber, B.~Blaettel, W.~Cassing, H.~C.~Doenges, V.~Koch, A.~Lang and U.~Mosel,
  Nucl.\ Phys.\ A {\bf 539}, 713 (1992).


\bibitem{Nara:2019qfd}
Y.~Nara and H.~Stoecker,
Phys. Rev. C \textbf{100}, no.5, 054902 (2019).

\bibitem{Nara:2020ztb}
Y.~Nara, T.~Maruyama and H.~Stoecker,
Phys. Rev. C \textbf{102}, no.2, 024913 (2020).







\bibitem{Danielewicz:1998pb}
  P.~Danielewicz, P.~B.~Gossiaux and R.~A.~Lacey,
  Fundam.\ Theor.\ Phys.\  {\bf 95}, 69 (1999)
  [nucl-th/9808013].

\bibitem{Sorensen:2020ygf}
A.~Sorensen and V.~Koch,
Phys. Rev. {\bf C} 104, 034904 (2021).

\bibitem{RLV}
  C.~Fuchs and H.~H.~Wolter,
  Nucl.\ Phys.\ A {\bf 589}, 732 (1995).

\bibitem{RQMD1995}
  H.~Sorge,
  Phys.\ Rev.\ C {\bf 52}, 3291 (1995).

\bibitem{UrQMD2}
  M.~Bleicher {\it et al.},
  J.\ Phys.\ G {\bf 25}, 1859 (1999).

\bibitem{JAMorg}
Y.~Nara, N.~Otuka, A.~Ohnishi, K.~Niita and S.~Chiba,
Phys.\ Rev.\ C {\bf 61}, 024901 (2000).

\bibitem{SMASH}
J.~Weil, V.~Steinberg, J.~Staudenmaier, L.~G.~Pang, D.~Oliinychenko, J.~Mohs, M.~Kretz, T.~Kehrenberg, A.~Goldschmidt and B.~B\"auchle, \textit{et al.}
Phys. Rev. C \textbf{94}, no.5, 054905 (2016).

\bibitem{jam2.1}
\url{https://gitlab.com/transportmodel/jam2}


\bibitem{Pythia8}
T.~Sj\"ostrand, S.~Ask, J.~R.~Christiansen, R.~Corke, N.~Desai, P.~Ilten, S.~Mrenna, S.~Prestel, C.~O.~Rasmussen and P.~Z.~Skands,
Comput. Phys. Commun. \textbf{191}, 159-177 (2015).
\url{https://pythia.org/}

\bibitem{Zhao:2020yvf}
X.~L.~Zhao, G.~L.~Ma, Y.~G.~Ma and Z.~W.~Lin,
Phys. Rev. C \textbf{102}, no.2, 024904 (2020).



\bibitem{Zhang:1999cd}
B.~Zhang, C.~M.~Ko, B.~A.~Li and A.~T.~Sustich,
J. Phys. G \textbf{26}, 1665-1670 (2000).


\bibitem{Li:2007yd}
Q.~Li, M.~Bleicher and H.~Stocker,
Phys. Lett. B \textbf{659}, 525-530 (2008).


\bibitem{Nayak:2019vtn}
K.~Nayak, S.~Shi, N.~Xu and Z.~W.~Lin,
Phys. Rev. C \textbf{100}, no.5, 054903 (2019).




\bibitem{Yong:2021npa}
G.~C.~Yong, Z.~G.~Xiao, Y.~Gao and Z.~W.~Lin,
Phys. Lett. B \textbf{820}, 136521 (2021).

\bibitem{Zhang:2021ddb}
D.~C.~Zhang, H.~G.~Cheng and Z.~Q.~Feng,
Chin. Phys. Lett. \textbf{38}, no.9, 092501 (2021).


\bibitem{Hartnack:1997ez}
C.~Hartnack, R.~K.~Puri, J.~Aichelin, J.~Konopka, S.~A.~Bass, H.~Stoecker and W.~Greiner,
Eur. Phys. J. A \textbf{1}, 151-169 (1998).









\bibitem{Rode:2018hlj}
S.~P.~Rode, P.~P.~Bhaduri, A.~Jaiswal and A.~Roy,
Phys. Rev. C \textbf{98}, no.2, 024907 (2018).


\bibitem{Huovinen:2001cy}
P.~Huovinen, P.~F.~Kolb, U.~W.~Heinz, P.~V.~Ruuskanen and S.~A.~Voloshin,
Phys. Lett. B \textbf{503}, 58-64 (2001).

\bibitem{STAR:2001ksn}
C.~Adler \textit{et al.} [STAR],
Phys. Rev. Lett. \textbf{87}, 182301 (2001).

\bibitem{Sun:2014rda}
X.~Sun, H.~Masui, A.~M.~Poskanzer and A.~Schmah,
Phys. Rev. C \textbf{91}, no.2, 024903 (2015)

\bibitem{Voloshin:1996nv}
S.~A.~Voloshin,
Phys. Rev. C \textbf{55}, R1630-R1632 (1997).


\bibitem{vanHecke:1998yu}
H.~van Hecke, H.~Sorge and N.~Xu,
Phys. Rev. Lett. \textbf{81}, 5764-5767 (1998).

\bibitem{Takeuchi:2015ana}
S.~Takeuchi, K.~Murase, T.~Hirano, P.~Huovinen and Y.~Nara,
Phys. Rev. C \textbf{92}, no.4, 044907 (2015).



\bibitem{Molnar:2003ff}
D.~Molnar and S.~A.~Voloshin,
Phys. Rev. Lett. \textbf{91}, 092301 (2003).

\bibitem{Lin:2003jy}
Z.~w.~Lin and D.~Molnar,
Phys. Rev. C \textbf{68}, 044901 (2003).

\bibitem{Fries:2003kq}
R.~J.~Fries, B.~Muller, C.~Nonaka and S.~A.~Bass,
Phys. Rev. C \textbf{68}, 044902 (2003).





\end{thebibliography}
\end{document}